\documentclass[twocolumn]{aastex631}
\usepackage{subfigure}
\usepackage{float}
\usepackage{pgfplots}
\usepackage{amsmath,bm}
\usepackage{amsfonts} 
\usepackage{amssymb}
\usepackage{booktabs}
\usepackage{multirow}
\usepackage[T1]{fontenc}
\usepackage{hyperref} % Load hyperref after all other packages to avoid conflicts

\begin{document}

%\title{High resolution  analysis of five exoplanets WASP-50\,b, WASP-117\,b, WASP-156\,b, WASP-167\,b and WASP-173A\,b}
\title{High-resolution spectroscopic atmospheric studies of 5 hot Jupiters across the edge of the Neptune desert}

\author[0000-0002-0486-5007]{Zewen Jiang}

\affiliation{State Key Laboratory of Particle Astrophysics, Institute of High Energy Physics, Chinese Academy of Sciences, Beijing 100049, China}

\author[0000-0002-9702-4441]{Wei Wang}\thanks{E-mail: wangw@nao.cas.cn}
\affiliation{National Astronomical Observatories, Chinese Academy of Sciences, Beijing 100101, China}
\affiliation{School of Astronomy and Space Science, University of Chinese Academy of Sciences, Beijing 100049, China}

\author[0000-0002-5203-8321]{Shumei Jia}
\affiliation{State Key Laboratory of Particle Astrophysics, Institute of High Energy Physics, Chinese Academy of Sciences, Beijing 100049, China}

\author[0009-0002-4524-3076]{Yaqing Shi}
\affiliation{Beijing Planetarium, Beijing academy of science and technology, Beijing 100044, China}

\author[0000-0001-9585-9034]{Fei Yan}
\affiliation{Department of Astronomy, University of Science and Technology of China, Hefei 230026, China}
\affiliation{Deep Space Exploration Laboratory, Hefei, Anhui 230026, China}

\author[0000-0003-2391-0093]{Li Zhou}
\affiliation{National Space Science Center, Chinese Academy of Sciences, Beijing 100190, China}

\author[0000-0001-5469-6443]{Qianyi Zou}
\affiliation{National Astronomical Observatories, Chinese Academy of Sciences, Beijing 100101, China}
\affiliation{School of Astronomy and Space Science, University of Chinese Academy of Sciences, Beijing 100049, China}

\author[0000-0002-2032-2440]{Xiang Ma}
\affiliation{State Key Laboratory of Particle Astrophysics, Institute of High Energy Physics, Chinese Academy of Sciences, Beijing 100049, China}

\author[0000-0002-3515-9500]{Yue Huang}
\affiliation{State Key Laboratory of Particle Astrophysics, Institute of High Energy Physics, Chinese Academy of Sciences, Beijing 100049, China}
%% Note that the \and command from previous versions of AASTeX is now
%% depreciated in this version as it is no longer necessary. AASTeX 
%% automatically takes care of all commas and "and"s between authors names.

%% AASTeX 6.31 has the new \collaboration and \nocollaboration commands to
%% provide the collaboration status of a group of authors. These commands 
%% can be used either before or after the list of corresponding authors. The
%% argument for \collaboration is the collaboration identifier. Authors are
%% encouraged to surround collaboration identifiers with ()s. The 
%% \nocollaboration command takes no argument and exists to indicate that
%% the nearby authors are not part of surrounding collaborations.

%% Mark off the abstract in the ``abstract'' environment. 
\begin{abstract} 
Hot Jupiters (HJs), especially the Ultra-Hot Jupiters (UHJs), are ideal targets for robust atmospheric characterization, thanks to their high equilibrium temperatures and large atmospheric scale heights, which result from their proximity to their host stars and intense stellar irradiation. Here, we present atmospheric studies of five planets, namely WASP-50b, WASP-117b, WASP-156b, WASP-167b, and WASP-173Ab. These five planets include two UHJs, two classic HJs, and one hot Neptune, with four of them just on the upper and middle borders of the Neptune desert, providing an interesting sample for investigating the connection between planetary atmospheric composition and bulk properties. We have not detected any significant absorption signals exceeding 3$\sigma$ in the three less-inflated, relatively high-density HJs (WASP-50b, WASP-156b, and WASP-173Ab). We marginally detect H$\alpha$ and Li {\sc i} with 3.2$\sigma$ and 3.1$\sigma$ in WASP-117b, respectively. In WASP-167b, we report tentative detection of H$\alpha$ and Fe {\sc i} at 4.6$\sigma$ and $\sim3.4\sigma$, receptively. In addition, Fe {\sc i} is significantly detected with a max SNR of 7.3 $\sigma$ using the cross-correlation technique, which exhibits a blue-shifted signal. For WASP-167\,b, we perform an atmospheric retrieval and yield the abundances of Fe, Mg, Ca, Ti, V, and equilibrium temperature of ${2479^{+193}_{-174}}$\,K. Comparing WASP-173A\,b and WASP-167\,b, both are UHJ, but with quite different extents of atmospheric signals, we propose that there may be a transition in $T_{\rm eq}$ between 1900 and 2300\,K.

\end{abstract}

%% Keywords should appear after the \end{abstract} command. 
%% The AAS Journals now uses Unified Astronomy Thesaurus concepts:
%% https://astrothesaurus.org
%% You will be asked to selected these concepts during the submission process
%% but this old "keyword" functionality is maintained in case authors want
%% to include these concepts in their preprints.
\keywords{Planets and satellites: atmospheres –- Planets and satellites: individual WASP-167b –- Methods: data analysis -- Techniques: spectroscopic}

%% From the front matter, we move on to the body of the paper.
%% Sections are demarcated by \section and \subsection, respectively.
%% Observe the use of the LaTeX \label
%% command after the \subsection to give a symbolic KEY to the
%% subsection for cross-referencing in a \ref command.
%% You can use LaTeX's \ref and \label commands to keep track of
%% cross-references to sections, equations, tables, and figures.
%% That way, if you change the order of any elements, LaTeX will
%% automatically renumber them.
%%
%% We recommend that authors also use the natbib \citep
%% and \citet commands to identify citations.  The citations are
%% tied to the reference list via symbolic KEYs. The KEY corresponds
%% to the KEY in the \bibitem in the reference list below. 

\section{Introduction} \label{sec:intro}
Exoplanet science has yielded significant insights and fundamentally altered our understanding of the universe since the discovery of the first exoplanet orbiting a Sun-like star~\citep{Mayor_1995}. The study of exoplanet atmospheres has emerged as a crucial research direction, providing valuable information on atmospheric properties, internal structures, formation processes, and evolutionary histories. Hot Jupiters (HJs), particularly Ultra-Hot Jupiters (UHJs) with equilibrium temperatures exceeding 2200 K~\citep{Kitzmann_2018,Parmentier_2018}, are characterized by their close proximity to their host stars and intense stellar irradiation. These factors result in high equilibrium temperatures and large atmospheric scale heights, which facilitate robust analyses of atmospheric components and enable repeated observations over short timescales, rendering UHJs ideal targets for detailed atmospheric characterization~\citep{Seager_2000,Brown_2001,Hubbard_2001,Seager_2010,Fossati_2018,Hoeijmakers_2018,Arcangeli_2018,Lothringer_2018}.

High-resolution spectroscopy (HRS) has emerged as one of the most powerful tools for characterizing exoplanet atmospheres~\citep{Snellen_2010,Wyttenbach_2015,Hoeijmakers_2018}. The increase in the number of spectral lines detectable by HRS enables sensitive detection of subtle shifts in these lines and variations in stellar spectra. This capability facilitates research into factors affecting planetary spectra, including exoplanet occultations, star spots, stellar pulsations, and atmospheric dynamics such as equatorial jet motions, atmospheric circulation, vertical convection, and atmospheric escape~\citep{Heng_2015}. Furthermore, HRS possesses exceptional capabilities for detecting species by inspecting the strong atomic lines in their rest frame and employing the cross-correlation function (CCF) method  for the densely packed and weak spectral lines that cannot be individually resolved. Numerous atoms and ions have been identified using HRS in various HJs and UHJs, including H~{\sc i}, He~{\sc i}, Li~{\sc i}, Na~{\sc i}, Mg~{\sc i}, K~{\sc i}, Ca~{\sc i}, Ti~{\sc i}, V~{\sc i}, Cr~{\sc i}, Mn~{\sc i}, Fe~{\sc i}, Co~{\sc i}, Ni~{\sc i}, Rb~{\sc i}, Sm~{\sc i}, Tb~{\sc i}, Ca~{\sc ii}, Sc~{\sc ii}, Ti~{\sc ii}, V~{\sc ii}, Fe~{\sc ii}, Co~{\sc ii}, Sr~{\sc ii}, and Ba~{\sc ii}~\citep{Wyttenbach_2015,Yan_2018,Nortmann_2018,Hoeijmakers_2019,Seidel_2019,Allart_2019,Hoeijmakers_2020,Chen_2020b,Tabernero_2021,Borsa_2021,Silva_2022,Kesseli_2022,Bello-Arufe_2022,Prinoth_2023,Jiang_2023a,Borsato_2023,Hoeijmakers_2024,DArpa_2024,Simonnin_2024,Prinoth_2025,Basinger_2025,Seidel_2025,Nortmann_2025}.

The ongoing advancements in data analysis techniques have significantly expanded the applicability of HRS. \citet{Brogi_2019} and \citet{Gibson_2020} establish a Bayesian framework for performing atmospheric retrieval based on HRS data, which enables the determination of species abundances, carbon-to-oxygen ratios (C/O), and metallicity. This framework provides insights into potential formation scenarios, giant planet migration mechanisms, and evolutionary pathways~\citep{Smith_2024,Pelletier_2025}. Furthermore, comparative studies with our own solar system enhance our understanding of the diversity of planetary systems.

In this paper, we present a comparative atmospheric study of WASP-50\,b, WASP-117\,b, WASP-156\,b, WASP-167\,b and WASP-173A\,b, 5 hot gas giant planets along the borders of the Neptune desert, in order to first obtain the knowledge of the atmospheres of them, and then try to compare the compositions of the planetary atmospheres and bulk parameters. The paper is organized as follows: Section~\ref{sec:observation} provides a comprehensive overview of these targets and the corresponding data-reduction processes. Section \ref{sec:methods} details the methodologies employed for data analysis, including fitting Rossiter-McLaughlin (RM) effects, generating transmission spectra, conducting cross-correlation analyses, and removing influences from RM and center-to-limb variation (CLV) effects, as well as the stellar pulsation, in addition to constraining species abundance in exoplanet atmospheres using the retrieval framework. Finally, Section \ref{sec:results and discussion} presents the results and subsequent discussions.

\section{SAMPLE and DATA}\label{sec:observation}
WASP-50\,b~\citep{Gillon_2011} is a HJ with a mass of 1.468 $M_{\rm J}$ and a radius of 1.153 $R_{\rm J}$. It has an equilibrium temperature of 1393 K and orbits a moderately bright G9 dwarf star with a period of 1.955 days. One transit of WASP-50\,b was observed on August 25, 2022, using the Echelle Spectrograph for Rocky Exoplanets and Stable Spectroscopic Observations (ESPRESSO) under the ESO program 0109.C-0319(G) (Principal Investigator: Simon Albrecht). ESPRESSO is a fiber-fed, ultra-stable echelle high-resolution spectrograph mounted at the 8.2 m Very Large Telescope at the European Southern Observatory in Cerro Paranal, Chile \citep{Pepe_2021}. This observation was conducted with a spectral resolving power of $R \sim 140000$ and a wavelength coverage of $380-788$ nm. During the transit of WASP-50\,b, a total of 61 spectra were obtained, with 25 collected in transit and 36 collected out of transit, covering an orbital phase $\Phi$ from $-0.046$ to 0.049. The mean signal-to-noise ratio (SNR) of the observed spectra at around 550 nm is approximately 39.

WASP-117\,b~\citep{Lendl_2014} is a HJ with a mass of 0.2755 $M_{\rm J}$ and a radius of 1.021 $R_{\rm J}$, featuring an equilibrium temperature of 1001 K. This exoplanet orbits a main-sequence F9 star with a period of 10.021 days, a period exceeding 10 days that was first identified by the Wide Angle Search for Planets (WASP)~\citep{Pollacco_2006}. One transit of WASP-117\,b was observed on October 26, 2018, using ESPRESSO under the ESO program 0102.C-0347(A) (PI: Fei Yan). For this transit observation, a total of 98 spectra were obtained, consisting of 62 spectra taken during the transit and 36 spectra taken outside of transit, covering an orbital phase $\Phi$ ranging from $-0.016$ to 0.023, with a mean SNR of approximately 48 around 550 nm.

WASP-156\,b~\citep{Demangeon_2018} is a super-Neptune located near the boundary of the Neptunian desert, with a mass of 0.128 $M_{\rm J}$ and a radius of 0.51 $R_{\rm J}$. The planet has an equilibrium temperature of 970\,K and orbits a K-type star with a period of 3.836 days. A transit of WASP-156\,b was observed on September 3, 2022, using the ESPRESSO spectrograph as part of the ESO program 0109.C-0745(C) (PI: Marina Lafarga Magro). During the transit, a total of 21 spectra were collected, comprising 12 spectra taken during the transit and 9 taken outside of transit, with orbital phase $\Phi$ ranging from $-0.028$ to 0.017. The mean SNR of the observed spectra around 550\,nm is approximately 48.

WASP-167\,b~\citep{Temple_2017} is a UHJ with a mass of 0.37 $M_{\rm J}$ and a radius of 1.621 $R_{\rm J}$ with equilibrium temperature of 2329\,K and an orbital period of 2.022 days, orbiting a rapidly rotating F1V star which is among the hottest stars known to host a transiting hot Jupiter. One transit of WASP-167\,b was observed on March 2, 2016, using the High Accuracy Radial Velocity Planet Searcher (HARPS) echelle spectrograph at the ESO 3.6 m telescope in La Silla, Chile. This spectrograph covers the optical range between 380 nm and 690 nm, with a spectral resolution of $R \sim 115000$~\citep{Mayor_2003}. Data were retrieved from the ESO archive under program 096.C-0762(A) (PI: Hellier, C). A total of 17 spectra were obtained, including 11 in-transit spectra and 6 out-of-transit spectra, with an average SNR of 41 around 550 nm, covering the phase $\Phi$ from $-0.052$ to 0.042.

WASP-173A\,b is a UHJ orbiting a Sun-like G2V star with an orbital period of 1.387 days and an equilibrium temperature of 1842\,K. It has a mass of 3.47\,$M_{\rm J}$ and a radius of 1.285\,$R_{\rm J}$. The host star is accompanied by a fainter companion, WASP-173B, which has a projected angular separation of approximately $\sim6$\arcsec~\citep{Labadie-Bartz_2019}. One transit of WASP-173A\,b was observed on July 24, 2022, using the ESPRESSO instrument as part of the ESO program 0109.C-0319(C) (PI: Albrecht Simon). During the transit of WASP-173A\,b, a total of 28 spectra were obtained, including 15 spectra taken in transit and 13 taken out of transit, with $\Phi$ ranging from $-0.069$ to 0.062. The mean SNR of the observed spectra at approximately 550\,nm is about 31.

\begin{figure}
	\includegraphics[width=\columnwidth]{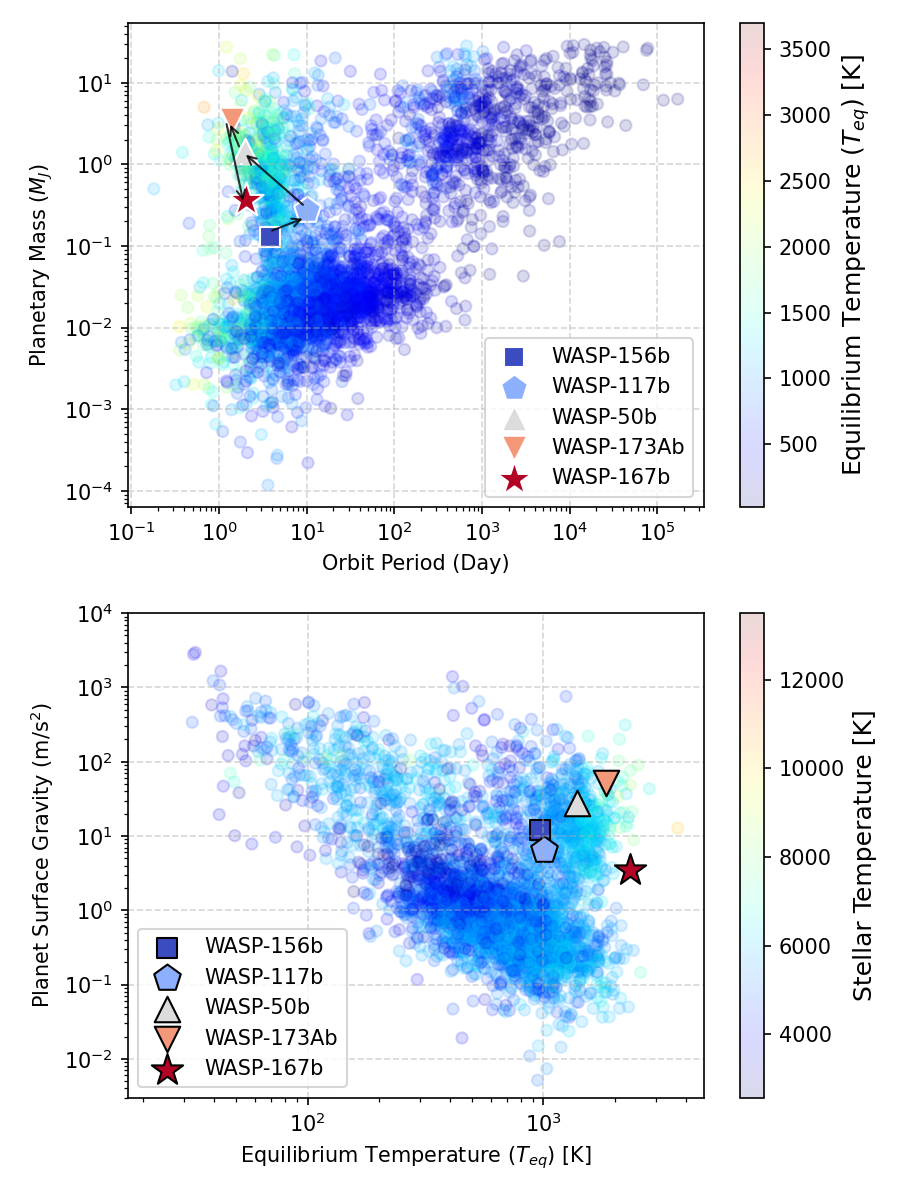}
    \caption{The distribution of all known planets in the planetary mass against orbit period diagram (top panel) and planetary surface gravity against planet equilibrium temperature ($T_{\rm eq}$) diagram (bottom panel), respectively. The five planets in this study are marked with different shapes and colors for clarity. Notably, $T_{\rm eq}$ of our sample planet spans from $\sim$1000\,K to $\sim$2300\,K. The arrows in the top panel represent the $T_{\rm eq}$ increasing direction.}
    \label{fig:PM_relationship}
\end{figure}

We show the sample considered in this study compared to all known exoplanets in Figure~\ref{fig:PM_relationship}. It is clear that WASP-173A\,b, WASP-50\,b and WASP-167\,b is around the upper border of the Neptune desert, while WASP-117b is in the middle border of the Neptune desert, and WASP-117b is away from the Neptune desert. Therefore, these 5 planets consist of an interesting although small sample that allow us to investigate the possible relationship between planetary atmospheres and bulk properties.

Standard echelle data reduction procedures were applied to the raw 2D spectra using the ESPRESSO Data Reduction Software (DRS) versions 2.2.1 and 2.5.0, as well as the HARPS DRS version 3.8. The one-dimensional (1D) spectra, which include the vacuum frame wavelengths, stellar fluxes, and associated uncertainties, can be extracted following the correction for bias, dark current, flat-fielding, and bad pixels. All details related to the observed information and planetary system parameters are summarized in Table~\ref{obs}.

% %%%%%%%%%%%%%%%%%%%%%%%%%%%%%%%%%%%%%%%%%%%%%
% \begin{figure}
% 	\includegraphics[width=\columnwidth]{P-M.png}
%     \caption{The distribution of candidate planets in the parameter space derived from the NASA Exoplanet Archive.}
%     \label{fig:PM_relationship}
% \end{figure}
% %%%%%%%%%%%%%%%%%%%%%%%%%%%%%%%%%%%%%%%%%%%%%%%
%%%%%%%%%%%%%%%%%%%%%%%%%%%%%%%%%%%%%%%%%%%%%
%%%%%%%%%%%%%%%%%%%%%%%%%%%%%%%%%%%%%%%%%%%%%%%

%-------------------------------------------------------------------%
\begin{table*}
  \centering
  \caption{Summary of the observations and planetary system parameters in this analysis.}
     \label{obs}
     \begin{tabular}{llllll}
        \hline
        \hline
        \noalign{\smallskip}
        Planets& WASP-50\,b &WASP-117\,b & WASP-156\,b & WASP-167\,b& WASP-173A\,b\\
        Instrument& ESPRESSO & ESPRESSO & ESPRESSO & HARPS &ESPRESSO \\
        Obs date& 2022-08-25 & 2018-10-26  & 2022-09-03 & 2016-03-02 &2022-07-24 \\
        Exp time& 240\,s & 300\,s & 700\,s & 1000\,s &555\,s \\
        Mean SNR& 38.7 & 48.1 & 47.87 & 40.61 &31.01 \\
        Mean Airmass& 1.28 & 1.35 & 1.2 & 1.09 &1.07 \\
        Mean Seeing& 0.77 & 1.0 & 0.84 & 0.83 &1.12 \\
        \hline
        Stellar Parameters&  &   &   &   &  \\
        \hline
        $m_{\rm v}$/mag&$11.44\pm0.026$ & $10.139\pm0.005$ & $11.559\pm0.019$ & $10.517\pm0.007$ &$11.102\pm0.034$\\
        $T_{\rm eff}$/K & $5400\pm100$ & $6040\pm90$  &$4910\pm61$  & $6900\pm150$ &  $5767_{-49}^{+50}$\\
        log\,$g_\star$/cgs& $4.537\pm0.022$ & $4.28\pm0.16$ & $4.40\pm0.12$ &$4.13\pm0.02$  & $4.393_{-0.060}^{+0.039}$\\
        $[\rm Fe/H]$/dex& -$0.12\pm0.08$ &-$0.11\pm0.14$  & $0.24\pm0.12$ & $0.1\pm0.1$ &  $5767_{-49}^{+50}$\\
        $M_\star$/$M_\odot$& $0.892_{-0.074}^{+0.080}$  &$1.126\pm$0.029   & $0.842\pm$0.052 &$1.49\pm0.13$  & $1.092_{-0.041}^{+0.045}$\\
        $R_\star$/$R_\odot$& $0.843\pm0.031$  & $1.170_{-0.059}^{+0.067}$ & $0.76_{-0.03}^{+0.03}$  &$1.861\pm0.057$  & $1.099_{-0.046}^{+0.079}$\\
        V\rm sin$i_{\star}$/$\rm {km\,s^{-1}}$& $2.6\pm0.5$ & $1.55\pm0.44$  & $3.80\pm0.91$  &$49.94\pm0.04^{2}$  & $7.9\pm0.5$\\
        \hline
        Planet Parameters&  &   &   &   &  \\
        \hline
        $M_{\rm p}$/$M_{\rm J}$& $1.468_{-0.086}^{+0.091}$ & $ 0.2755\pm0.009$ &$0.128_{-0.009}^{+0.010}$   & $ 0.37\pm0.22$&$3.47_{-0.14}^{+0.15}$\\
        $R_{\rm p}$/$R_{\rm J}$& $1.153\pm0.048$ &$ 1.021_{-0.065}^{+0.076}$  &  $ 0.51_{-0.02}^{+0.02}$&
        $ 1.621\pm0.059$ &$ 1.285_{-0.071}^{+0.12}$\\
        $\rho$/g\,cm$^{-3}$ &  $0.958_{-0.082}^{+0.095}$ &$0.259_{-0.048}^{+0.054}$  &$1.0\pm0.1$  &$0.116\pm0.070$ &$2.02_{-0.47}^{+0.38}$\\
        $T_{\rm equ}$/K& $1393\pm30$   &$1001_{-32}^{+29}$ &  $970_{-20}^{+30}$&$2329\pm64$ &$1842_{-42}^{+65}$\\
        $R_{\rm p}/R_\star$ & $0.1404\pm0.0013$  &$0.08611_{-0.00076}^{+0.00048}$  &$0.0685_{-0.0012}^{+0.0008}$  &$0.08945\pm0.00051$ &$0.1203_{-0.0028}^{+0.0032}$\\
        \hline
        Orbital Parameters&  &   &   &   &  \\
        \hline
        $T_{\rm c}$/BJD$_{\rm TDB}$&$5859.6968\pm0.00012^{1}$  &$6533.824038_{-0.00090}^{+0.00095}$  &$4677.707_{-0.002}^{+0.002}$  &$8592.18253\pm0.00027$ &$8105.59824_{-0.00090}^{+0.00084}$\\
        $P$/day& $1.9550959\pm5.1\rm e^{-6} $   & $10.0205931\pm5.5 \rm e^{-4} $  &  $3.836169\pm3.0\rm e^{-6} $&$2.02195830\pm5.0\rm e^{-7} $ &$1.3866529\pm2.7\rm e^{-6} $\\
        $T_{14}$/day& $0.07524\pm0.00068 $   &  $0.2475_{-0.0029}^{+0.0033}$ & $0.10042_{-0.00125}^{+0.00167}$  & $0.1135\pm0.0008^{2} $ &$0.0981_{-0.0022}^{+0.0025}$\\
        $a$/AU& $0.02945\pm0.00085$  & $0.09459_{-0.00079}^{+0.00084}$ &$0.0453_{-0.0009}^{+0.0009}$  &$0.0365\pm0.0006^{2}$ &$0.02508_{-0.00032}^{+0.00034}$\\
        $e$& $0.009_{-0.006}^{+0.011}$  & $0.302\pm0.023$ &$<0.007$  &- &$0$\\
        $\omega$/deg& $44_{-80}^{+62}$  & $242_{-2.7}^{+2.3}$ & - &- &-\\
        $i$/deg&   $84.74\pm0.24$ &$89.14\pm0.30$  &$89.1_{-0.9}^{+0.6}$  &$79.3\pm0.2$ &$86.5_{-2.3}^{+2.2}$\\
        $K_{\rm p}$/\rm km$\,\rm s^{-1}$&   $256.6\pm 4.4$ & $25.16\pm 0.69$ &$19\pm 1$  &$45\pm 34$ &$592\pm 20$\\
        \noalign{\smallskip}
        \hline
        \hline
        \noalign{\smallskip}
     \end{tabular}
     
     \textbf{Notes}:The main references for WASP-50\,b in this paper is~\citet{Gillon_2011}, (1):~\citet{Tregloan-Reed_2013}. WASP-117\,b is~\citet{Lendl_2014,Bonomo_2017,Kokori_2023,Saha_2024}. WASP-156\,b is ~\citet{Demangeon_2018,Saha_2021}. WASP-167\,b is ~\citet{kalman_2024}, (2):~\citet{Temple_2017}. WASP-173A\,b is ~\citet{Labadie-Bartz_2019,Knudstrup_2024,Zak_2025}. The value of $T_{\rm c}$ is subtracted by 2,450,000 in table.
 \end{table*}

 %-----------------------------------------------------------------%
\section{METHODS}\label{sec:methods}
\subsection{Rossiter-McLaughlin analysis}
\label{sec:rm analysis}
The measured radial velocity (RV) curve, anticipated for the Keplerian orbit of the targeted star, may deviate when a planet transits in front of its rotating host star. This phenomenon is known as the Rossiter-McLaughlin (RM) effect, which was first identified by~\citet{Rossiter_1924} and ~\citet{McLaughlin_1924} while studying the eclipse effect reflected in the velocity space of binary stars. The same effect was first observed in a planetary system by~\citet{Queloz_2000}, which detected the distortion of the stellar line profiles during a planetary transit. Analyzing the RM effects can provide valuable information about the planet's orbital behavior over time and offer more precise constraints on planetary formation theories~\citep{Dawson_2018}.

To analyze and model the RM effect, we employed the Markov Chain Monte Carlo (MCMC) algorithm implemented in the \texttt{emcee} package~\citep{Foreman-Mackey_2013}. Additionally, we utilized the RM effect model integrated in the \texttt{rmfit} code\footnote{\url{https://github.com/gummiks/rmfit}} to generate the RM model, following the methodologies described in ~\citet{Stefansson_2022} and ~\citet{Hirano_2011}. We placed Gaussian priors on the transit center time ($T_{0}$), planet-to-star radius ratio ($R_{p}/R_{\star}$), scaled semi-major axis ($a/R_{\star}$), limb-darkening coefficient ($u$), planetary orbital inclination ($i$), semi-amplitude of radial velocity ($K$), eccentricity ($e$), argument of periastron ($\omega$), and uniform priors on sky-projected spin-orbit angle between the orbital plane and the apparent equatorial plane ($\lambda$), intrinsic stellar line width ($\beta$), sky-projected rotational velocity of the host star ($V\sin i_{\star}$), systematic velocity ($\gamma$) and jitter term ($\sigma_{jitter}$). The remaining parameters, including the orbital period $P$, were held constant at values reported in the literature, as detailed in Table~\ref{obs}. It also should be noted that $e$ and $\omega$ are not taken into account but fixed to be 0 and 90 degrees respectively when modeling the RM effects given their short orbital periods for WASP-156b and WASP-173Ab. In this study, we employed the CCF method within the ESPRESSO pipeline to derive the values of radial velocity (RV) and associated uncertainties. We focused our analysis on the RM effect of WASP-117, WASP-156, and WASP-173A using the RV data to achieve more precise orbital parameters. To improve data quality, we discarded the last three RV data points of WASP-173A due to substantial variations compared to other data, and the remaining data were binned every two adjacent points. The relevant parameters for WASP-50 and WASP-167 were directly adopted as presented in \citet{Knudstrup_2024} and \citet{Temple_2017}.

To establish a comprehensive constraint on the distribution of selected parameters, we employed a configuration of 100 walkers, each executing 10,000 steps. The initial 5,000 steps were excluded as burn-in, facilitating an effective exploration of the parameter space and enabling the probability density to stabilize within the region of maximum likelihood during the MCMC analysis. The chosen priors and the resulting posteriors for each parameter derived from the MCMC fitting are presented in Table~\ref{prior}. Based on the analysis of the results of fitting RM effects, WASP-117\,b is in a nearly aligned orbit with $\lambda = 27.307^{+4.528}_{-4.299}$ deg. Additionally, we also performed an RM fit using the HARPS RV data provided by \citet{Lendl_2014} with the same priors, which results in a $\lambda$ of $34.199^{+15.335}_{-18.801}$ deg. The resulting $\lambda$ values are essentially consistent within their $1\sigma$ errors, but they show a significant discrepancy relative to the values of $-44\pm 11$ deg reported by \citet{Lendl_2014}. We adopted the $\lambda$ derived by us, given the higher SNR of the ESPRESSO data. WASP-156\,b is likely to be in a well-aligned orbit, as indicated by the derived small but uncertain $\lambda = -5.15^{+34.572}_{-33.477}$\,deg, orbiting a slowly rotating star with $V\sin i_{\star} = 0.639^{+0.329}_{-0.16}$ km\,s$^{-1}$. Future more extensive and precise RV data may provide better constraints on $\lambda$; WASP-173A\,b also has an aligned orbit with $\lambda = 8.8877^{+3.6842}_{-3.7735}$ deg, which is nearly consistent with the value $\lambda = 11^{+32}_{-20}$ deg reported by \citet{Knudstrup_2024} using the same ESPRESSO datasets, and also orbits its host star with $V\sin i_{\star} = 7.717^{+0.367}_{-0.34}$ km\,s$^{-1}$, a similar and reasonable uncertainty comparable to the 6.1\,$\pm\,0.3$ km\,s$^{-1}$ in \citet{Hellier_2019} and 7.9\,$\pm\,0.5$ km\,s$^{-1}$ in  \citet{Labadie-Bartz_2019}. We adopted these values, as shown in Table~\ref{obs}, as input parameters to model and correct for the center-to-limb variation (CLV) effect in combination with the RM effect, which mitigates their impact on the
final transmission spectra and a more precise detection of atmospheric composition, enhancing the accuracy and reliability of the derived results.
%---------------------Figure:RV FIT--------%
\begin{figure*}
    \centering
    \subfigure[]{\includegraphics[width=0.23\textwidth]{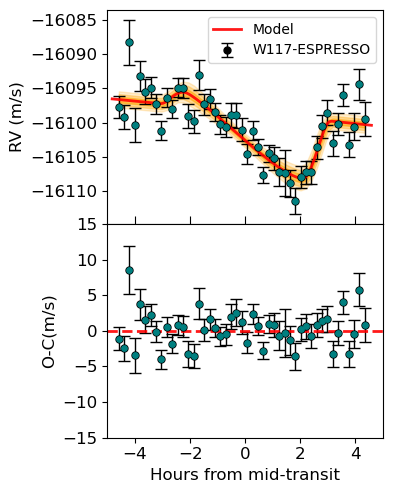}}
    \subfigure[]{\includegraphics[width=0.23\textwidth]{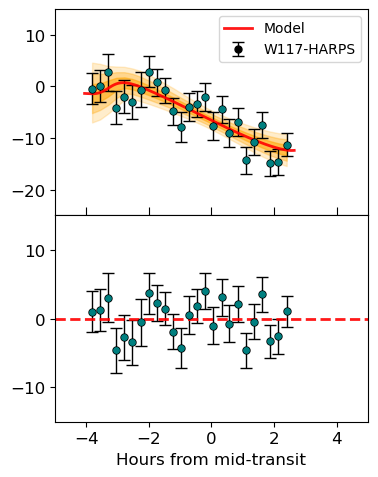}}
    \subfigure[]{\includegraphics[width=0.23\textwidth]{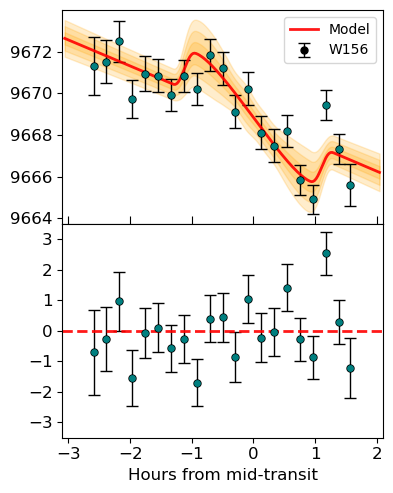}}
    \subfigure[]{\includegraphics[width=0.23\textwidth]{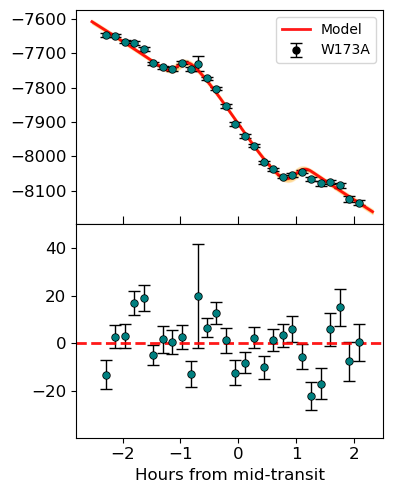}}
    \caption{{\textit{Top panel}:} The RV curve produced by the spectrograph instrument for different targets: WASP-117 (a,b), WASP-156 (c), and WASP-173A\,(d), showing the RM effect and Kepler motion with green solid circles and error bars. The best-fit model using the \texttt{rmfit} procedure is shown in the solid red line, the orange shaded region indicates the 1$\sigma$, 2$\sigma$ and 3$\sigma$ models. \textit{Bottom panel}: The residuals after removing the model prediction from the RV data.}
    \label{fig:RM fit}
\end{figure*}

%---------------------Figure:RV FIT--------%

%-------------------------------------table--------%
\begin{table*}
\small
  \caption[]{Summary of priors and resulting posteriors derived from the RV curve fitting.}
     \label{prior}
     $$
     \begin{tabular}{p{0.10\linewidth}llllllll}
        \hline
        \hline
        \noalign{\smallskip}
         Parameter&\multicolumn{3}{c}{WASP-117\,b} & \multicolumn{2}{c}{WASP-156\,b} &\multicolumn{2}{c}{WASP-173A\,b} \\ \cline{2-4} \cline{5-6}\cline{7-8}         
         &Prior & Posterior$^{2}$ & Posterior$^{3}$ &  Prior & Posterior & Prior & Posterior \\
        \hline
        \noalign{\smallskip}
        $T0^{1}$/BJD& N(0,\,0.001) &$-0.0001_{-0.001}^{+0.001}$&$0_{-0.001}^{+0.001}$&N(0,\,0.001)&$0.0003_{-0.0009}^{+0.0009}$&N(0,\,0.001)&$0.0011_{-0.0005}^{+0.0005}$\\  
        \noalign{\smallskip}
        $R_{p}/R_{\star}$ &N(0.0861,\,5e$^{-4}$)&$0.0861_{-0.0005}^{+0.0005}$&$0.0861_{-0.0005}^{+0.0005}$&N(0.0676,\,0.001)&$0.0675_{-0.001}^{+0.001}$&N(0.12,\,0.003)&$0.1268_{-0.0025}^{+0.0025}$ \\
        \noalign{\smallskip}
         $a/R_{\star}$ & N(13.12,\,0.25) &$13.486_{-0.21}^{+0.22}$&$13.114_{-0.253}^{+0.258}$&N(12.748,\,0.025)&$12.749_{-0.0262}^{+0.0247}$&N(4.85,\,0.14)&$4.9607_{-0.0719}^{+0.066}$\\  
        \noalign{\smallskip}
         $u1$ & N(0.327,\,0.06)&$0.334_{-0.061}^{+0.059}$&$0.327_{-0.06}^{+0.06}$&N(0.472,\,0.004)&$0.472_{-0.004}^{+0.004}$&N(0.4,\,0.01)&$0.4021_{-0.0102}^{+0.0101}$\\
        \noalign{\smallskip}
         $u2$ &N(0.148,\,0.01)&$0.148_{-0.01}^{+0.01}$&$0.148_{-0.01}^{+0.01}$&N(0.134,\,0.007)&$0.134_{-0.007}^{+0.007}$&N(0.3,\,0.01)&$0.3015_{-0.01}^{+0.01}$\\  
        \noalign{\smallskip}
        $i$/deg& N(88.15,\,0.2) &$87.574_{-0.105}^{+0.108}$&$88.14_{-0.223}^{+0.221}$&N(89.1,\,0.6)&$88.926_{-0.694}^{+0.736}$&N(86.5,\,0.1)&$86.5057_{-0.0972}^{+0.0967}$\\  
        \noalign{\smallskip}   
        $K$/m s$^{-1}$ & N(25.16,\,0.69) &$25.08_{-0.68}^{+0.67}$&$25.16_{-0.70}^{+0.67}$&N(19,\,1)&$18.566_{-0.934}^{+0.951}$&N(612,\,5)&$626.345_{-3.5494}^{+3.5398}$\\
        \noalign{\smallskip}
        $e$/deg& N(0.302,\,0.023) &$0.296_{-0.023}^{+0.022}$&$0.302_{-0.023}^{+0.024}$& -&-&-&-\\ 
        \noalign{\smallskip}
        $\omega$/deg& N(242,\,2.3) &$242_{-2.28}^{+2.31}$&$241.870_{-2.273}^{+2.318}$& -&-&-&-\\
        \noalign{\smallskip}
        $\lambda$/deg& U(-180,\,180) &$27.31_{-4.30}^{+4.53}$&$34.199_{-18.801}^{+15.335}$& U(-180,\,180)&$-5.15_{-33.477}^{+34.572}$&U(-180,\,180)&$8.8877_{-3.7735}^{+3.6842}$\\
        \noalign{\smallskip}
        $\beta$/km s$^{-1}$& U(0.5,\,30) &$16.64_{-9.25}^{+9.15}$&$16.68_{-9.75}^{+9.03}$&U(0.5,\,30)&$13.516_{-9.623}^{+11.291}$&U(0.5,\,30)&$16.6291_{-5.7274}^{+8.4169}$\\  
        \noalign{\smallskip}
        V\rm sin$i_{\star}$/km s$^{-1}$& U(0.1,\,10) &$1.886_{-0.175}^{+0.188}$&$1.697_{-0.315}^{+0.386}$&U(0.1,\,10)&$0.6387_{-0.16}^{+0.3291}$&N(6.2,\,0.5)&$7.7165_{-0.3402}^{+0.3671}$\\ 
        \noalign{\smallskip}
        $\gamma$/km s$^{-1}$& U(-50,\,5) &$-16.095_{-0.001}^{+0.001}$&$0.001_{-0.002}^{+0.002}$&U(0,\,20)&$9.6688_{-0.0002}^{+0.0002}$&U(-20,\,0)&$-7.8924_{-0.0016}^{+0.0016}$\\  
        \noalign{\smallskip}
        ln\,$\sigma_{\rm jitter}$/km s$^{-1}$& U(-6,\,0) &$-3.08_{-1.98}^{+2.09}$&$-3.05_{-2.01}^{+2.03}$&U(-6,\,0)&$-2.966_{-2.038}^{+1.967}$&U(-6,\,0)&$-3.0408_{-2.019}^{+2.0807}$\\ 
        \hline
     \end{tabular}
     $$
     \textbf{Notes}:\textit{U(a,b)} represents a uniform distribution with a low and high limit of a and b, respectively; \textit{N(m,$\sigma$)} denotes a gaussian distribution with a mean value and standard deviation of m and $\sigma$. $T0$: Transit center time, $R_{p}/R_{\star}$: Planet-to-star radius ratio, $a/R_{\star}$: Scaled semi-major axis, $u1$: Linear limb dark coefficient, $u2$: Quadratic limb dark coefficient, $i$: Orbital inclination, $K$: Semi-amplitude of RV, $e$: Eccentricity, $\omega$:Argument of periastron, $\lambda$: Projected spin-orbit angle, $\beta$: Intrinsic stellar line width, V\rm sin$i_{\star}$: Projected rotational velocity, $\gamma$: Systematic velocity, ln\,$\sigma_{\rm jitter}$: Jitter term for instrument. (1): Each $T0$ minus the $T_{c}$ value. (2): The posterior calculated by ESPRESSO data. (3): The posterior calculated by HARPS data but remove the baseline of Keplerian orbital velocity.
\end{table*}
%-------------------------------------------------%

\subsection{Transmission spectrum construction}
\label{sec:Transmission_spectra}

We investigated the presence of atmospheric species on five exoplanets using the methodology established by \citet{Wyttenbach_2015} and \citet{Casasayas-Barris_2019}. First, we employed the ESO \texttt{Molecfit} software, version 1.5.7~\citep{Smette_2015,Kausch_2015}, to correct for the telluric absorption introduced by Earth's atmosphere, which is superimposed on the stellar spectra. We then constructed the master-out spectrum by calculating the average of the out-of-transit spectra, weighting each exposure by its SNR after normalizing and aligning each spectrum to the stellar rest frame~\citep[e.g.,][]{Stangret_2021}. The master-out spectrum is should be purely stellar origin, and is thus used to divide each individual spectrum in order to remove stellar light contribution so that the absorption signal from the planetary atmosphere may emerge in residual spectrum. The final one-dimensional (1D) transmission spectra were obtained by combining the in-transit residuals, which were shifted into the planet rest frame.

\begin{figure}
	% To include a figure from a file named example.*
	% Allowable file formats are eps or ps if compiling using latex
	% or pdf, png, jpg if compiling using pdflatex
	\includegraphics[width=\columnwidth]{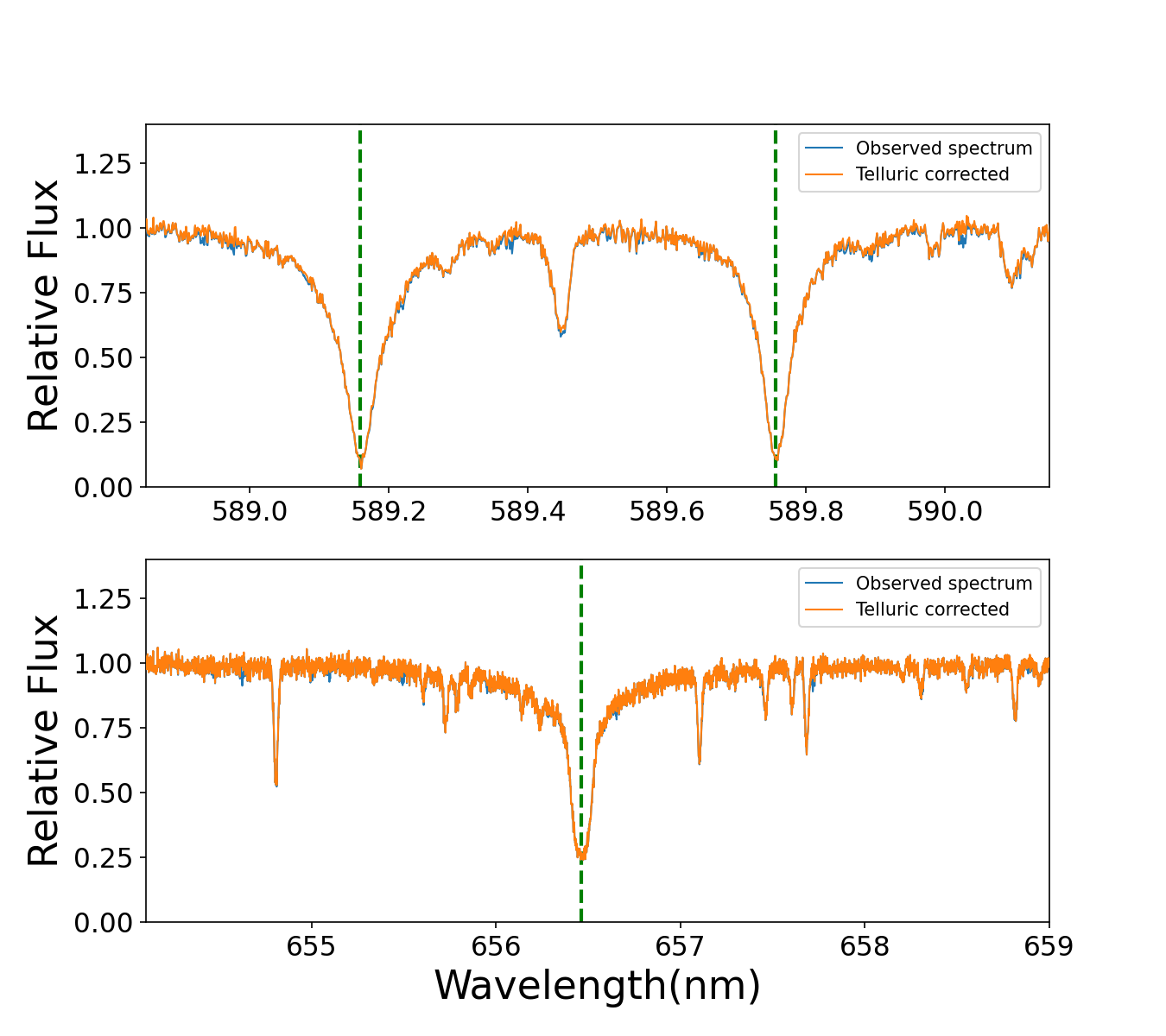}
    \caption{An example of telluric correction using \texttt{Molecfit} in the observed spectrum of WASP-173A\,b for Na~{\sc i} D1 $\&$ D2 (\textit{Top}) and H~{\sc i} (\textit{Bottom}). The observed spectra are shown in blue, while the spectra after telluric correction are shown in orange. The dashed green vertical lines represent the static positions of the atomic lines in vacuum.}
    \label{fig:telluric_correction_figure}
\end{figure}

It is worth noting that an elliptical orbit was adopted specifically for WASP-117b to account for its significant $e$ and $\omega$ as shown in Table~\ref{obs}, we employed the RVmodel algorithm implemented in the \texttt{RadVel} package~\citep{Fulton_2018} to calculate the stellar RV value at different orbital phase and correct the velocity for planetary rest frame. The circular orbits are adopted for the other planets as their short periods. Additionally, the observed and contamination-corrected spectra of WASP-50 around the Na D1$\&$D2 doublet region (top panel) and H$\alpha$ spectral line region (bottom panel) are shown in Fig.~\ref{fig:telluric_correction_figure}, illustrating the effectiveness of the telluric correction. We also note that there is a wiggle pattern in the derived transmission spectrum that only identified in ESPRESSO data, which might potentially affect the final spectral signature analysis and need to be corrected.\,These wiggles were first reported in \citet{Allart_2020} and are possibly induced by an interference pattern caused by the Coudé train optics as pointed out by \citet{Tabernero_2021}. To correct these wiggles, we calculate the median values of the intervals by binning the transmission spectra using several selected step sizes and fit the global fluctuations of this median dataset to remove the sinusoidal trend in the obtained transmission spectra.

The deformation of stellar line profiles produces spurious signals in the final transmission spectra, which result from center-to-limb variations~\citep{Cegla_2016,Yanfei_2017} and the Rossiter-McLaughlin effect ~\citep[RM,][]{Rossiter_1924, McLaughlin_1924}. These phenomena arise due to the non-uniform brightness and variations in Doppler effects across the stellar discs. To mitigate these influences on the planetary signal, we adopted the methodology utilized by~\citet{Yan_2018}, \citet{Casasayas_Barris_2019}, and \citet{Chen_2020a} to model the stellar spectra at different transit positions. We employed the \texttt{Spectroscopy Made Easy}~\citep[SME,][]{Valenti_1996} tool to calculate the theoretical stellar spectra at 21 distinct limb-darkening angles ($\mu$) using the MARCS and VALD3 line lists~\citep{Ryabchikova_2015}, while adopting corresponding abundance and local thermodynamic equilibrium (LTE). Additionally, the stellar disk is also divided into small elements with size of $0.01\,R_\star\times0.01\,R_\star$, and each of them owns specific parameters including $v\,{\rm sin}\, i_{\star}$, $\mu$, and $\theta$. Then the synthetic spectrum at different orbital phase is calculated by summing up the spectra of the stellar disk, excluding those elements obscured by the planet.

\begin{figure}
    \includegraphics[width=8.0cm, height=7cm]{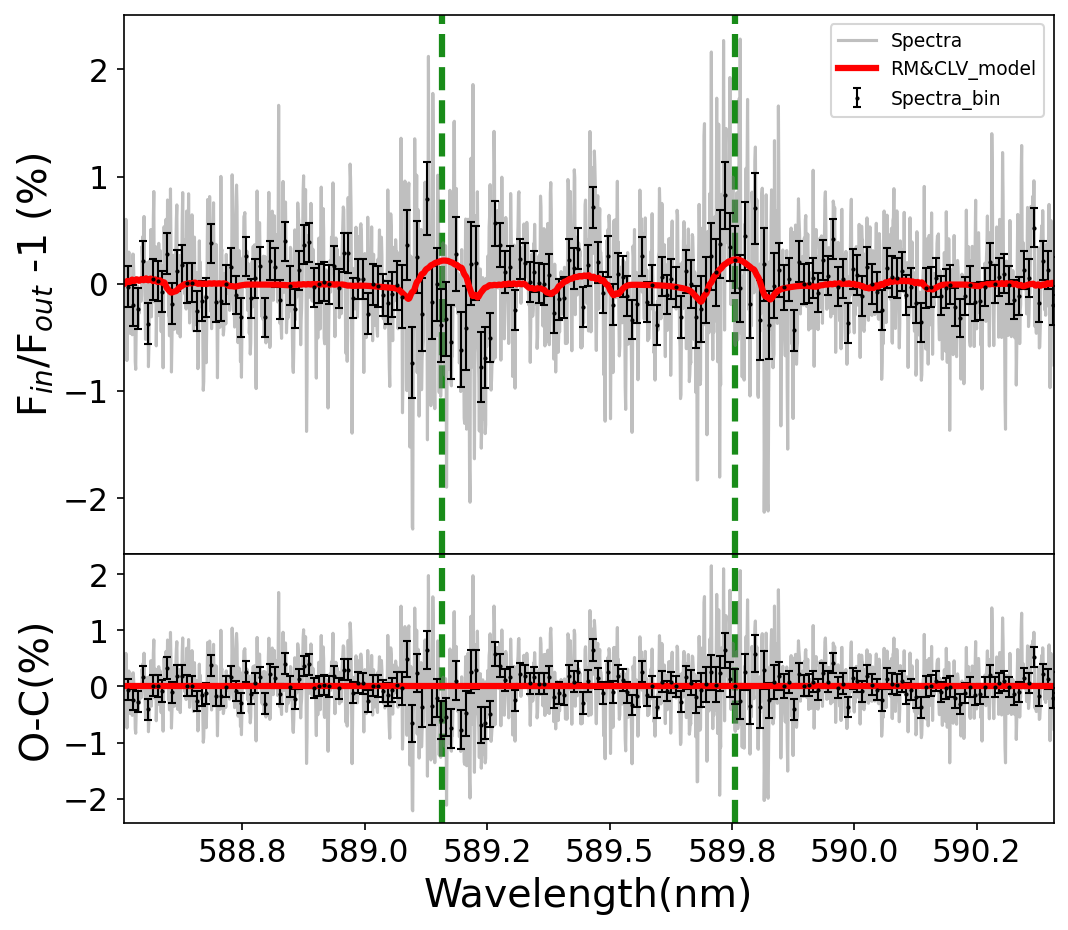}
    \caption{{\textit{Top panel}:} 
    The model of RM and CLV effects overplotted in solid red lines on the final transmission spectra which is shown in solid grey lines as background around the Na doublet lines, the binned transmission spectrum with error bar and a size of 0.1\textup{~\AA} is shown in black points. {\textit{Bottom panel}: The residuals after removing the model of RM and CLV effects.} 
    }
    \label{fig:1d_rm_clv_model}
\end{figure}

The subsequent step involves generating and correcting the RM+CLV model. We construct the master out-of-transit spectrum using the synthetic spectrum obtained from \texttt{SME}. Each synthetic spectrum during transit is divided by this master spectrum, and the RM+CLV effect model is derived by summing the residuals and rescaling to match the transmission spectrum. After subtracting the RM+CLV effect, detectable signals will manifest in the 1D transmission spectrum. Figure~\ref{fig:1d_rm_clv_model} illustrates the CLV and RM effects in the residual spectra of the sodium doublets for WASP-173A\,b, highlighting the effectiveness of eliminating both effects from the transmission spectra. Nonetheless, the correction may not be optimal for certain specific wavelength regions and planets.

Finally, In order to further confirm the existence of these atmospheric signals, the technique of empirical Monte Carlo (EMC) simulations~\citep{Redfield_2008} is performed to estimate the effect of systematics, which has been widely used in previous studies~\citep{Wyttenbach_2015, Seidel_2019, Allart_2020, Casasayas_2021}. The concept of this method is to artificially create new data set by randomizing original data to verify whether the investigated signals still exist. We explored three scenarios including in-in, in-out, and out-out for each strong line, with 10000 iterations for the $\sim$1.5\,{\AA}\ passband, and we adopted a sample size with 50\% of the number of in or out-of-transit data to create new realization. The detailed results will be presented and discussed in section~\ref{sec:results and discussion}.

%%%%%%%%%%%%%%%%%%%%%%%%%%%%%%%%%%%%%%%%%%%%%
% First figure
%%%%%%%%%%%%%%%%%%%%%%%%%%%%%%%%%%%%%%%%%%%%%%%

%%%%%%%%%%%%%%%%%%%%%%%%%%%%%%%%%%%%%%%%%%%%%%%%%%%%%
%%%%%%%%%%%%%%%%%%%%%%%%%%%%%%%%%%%%%%%%%%%%%%%%%%%%%

\subsection{Cross-correlation analysis}
\label{sec:ccf}
To enhance the identification of potential atmospheric signals characterized by multiple spectral lines in these exoplanets, we conducted a comprehensive survey of species from lithium (Li) to uranium (U) utilizing the cross-correlation function technique (CCF) to compare the obtained transmission spectra with atmospheric template spectra. The template spectra were generated using the Python package \texttt{petitRADTRANS}~\citep{P_Molliere_2019}, which has been effectively employed in various atmospheric studies~\citep{PMolliere_2020, Cont_2024}. To address gaps in opacity within \texttt{petitRADTRANS}, we incorporated opacity data from the Data and Analysis Center for Exoplanets~\citep[DACE\footnote{\url{https://dace.unige.ch/opacityDatabase/}},][]{Grimm_2015, Grimm_2021}, which includes data from the Kurucz~\citep{Kurucz_2018}, National Institute of Standards and Technology~\citep[NIST,][]{NIST_ASD}, and Vienna Atomic Line Data Base ~\citep[VALD3,][]{Ryabchikova_2015} databases. The opacity for all the elements included on DACE database were entirely adopted here. The template spectra also require some physical parameters related to the planetary atmosphere as input. Here, we adopted a Guillot temperature profile calculated by the atmospheric pressure, opacity in optical and infrared wavelengths, planetary equilibrium temperature and the planetary internal temperature. Free Chemistry and hydrostatic equilibrium is also assumed for each planet to generate the atmospheric template. Additionally, we adjusted the factors of the line-by-line opacities to align with the resolution of spectra obtained using the ESPRESSO and HARPS instruments.

%----------------------------------%
\begin{equation} \label{ccf_formula}
\centering
	c(v,t)= \frac{\sum_i^N x_i(t) T_i(v)}{\sum_i^N T_i(v)}
\end{equation}
%----------------------------------%
where $c(v,t)$ represents a two dimensional matrix that is dependent on both $t$ and $v$, $i$ is the wavelength index, $x_i(t)$ refers to the transmission spectrum at a specific $t$, while $T_i(v)$ denotes the template that has been shifted to a radial velocity of $v$.

If the targeted species exists, the signal will appear as a dark trace along the expected moving direction of the planet in the CCF map, and prominently manifest at the position of the estimated orbital velocity $K_{\rm p}$ and system velocity $V_{\rm sys}$ in the $K_{\rm p}-\Delta V_{\rm sys}$ map.

\subsection{Removal of RM \& CLV effects and stellar pulsations}
\begin{figure}
	\includegraphics[width=8.5cm, height=5cm]{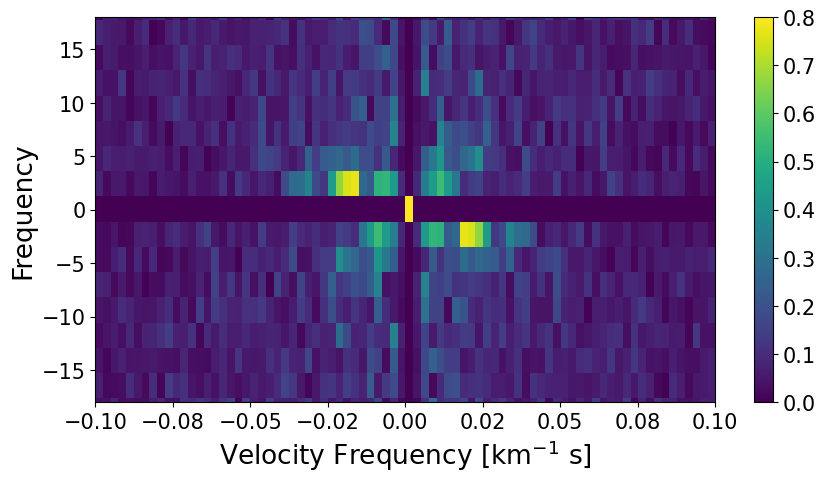}
    \caption{
    The Fourier transform of result after performing cross-correlation with Fe for WASP-167\,b. The CLV and RM effects are characterized by the feature from the bottom-left to the top-right, the stellar pulsations are seen as the feature aligned with horizontal axis extending from the bottom-right to the top-left.}
    \label{fig:fft}
\end{figure}

As shown in the Figure~\ref{fig:fft}, the top panel delineates the result of WASP-167\,b obtained from the CCF analysis with template spectra only containing iron element. The result mainly reveals two distinct structural components, one is the stellar pulsation signal characterized by the alternating bright yellow and dark striations from bottom left to top right throughout the entire transit duration, the other is the bright yellow strip extending from bottom right to top left, which is attributed to the CLV and RM effects. These two structures can produce strong deformation on the stellar profiles and significantly diminish the detectability of planetary signal along the trace of orbital motion.

\begin{figure}
	\includegraphics[width=8cm, height=8cm]{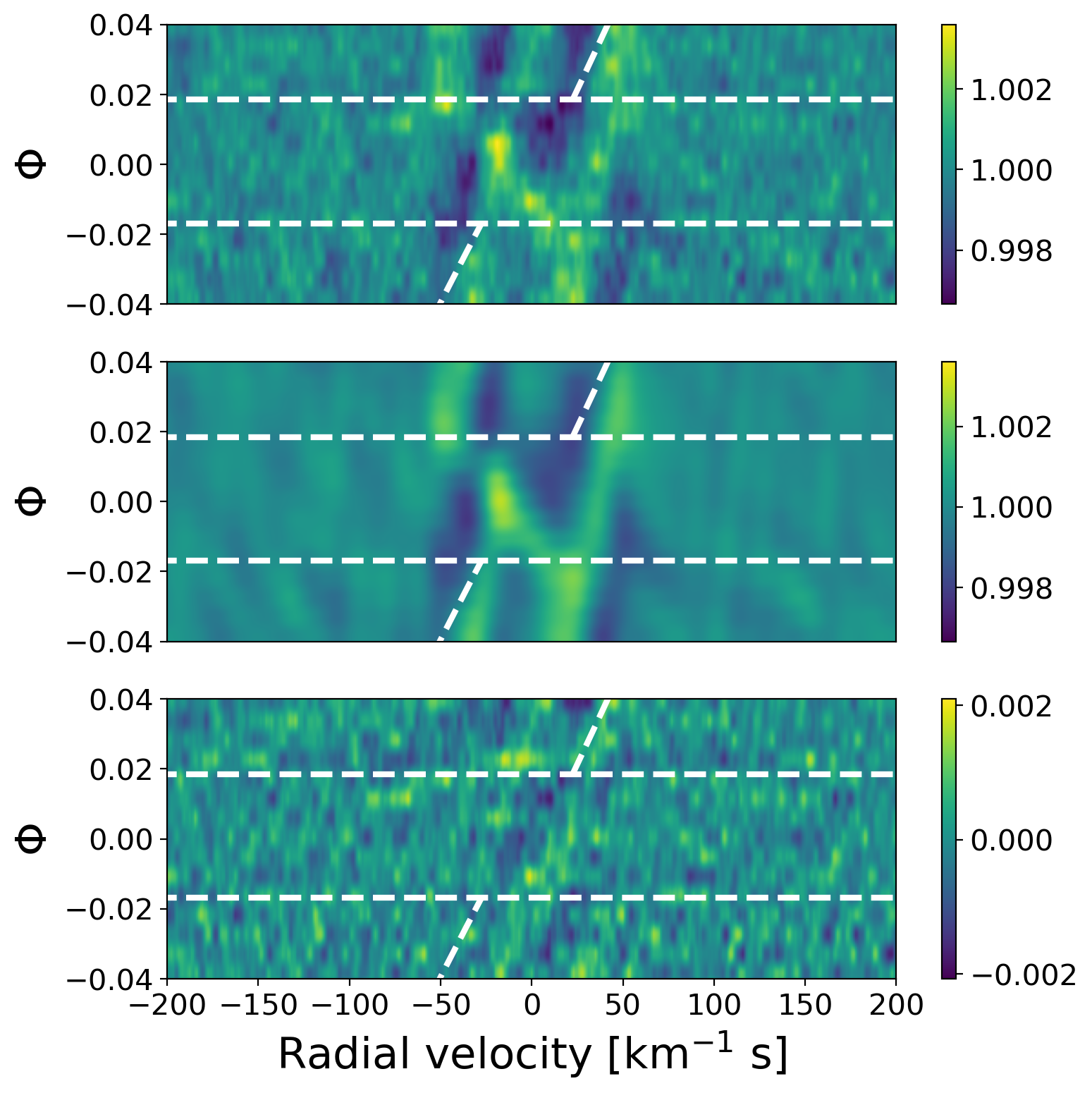}
    \caption{The CCF Map of WASP-167b at different orbital phase (\textit{top panel}), the extracted model after the application of Fourier transform (\textit{middle panel}) and the residual after subtracting the model (\textit{bottom panel}), respectively. In each panel, the white dotted lines mark the beginning and ending contacts during the transit, the inclined white lines indicate the expected trace of signal from the planetary atmosphere, and the alternating bright yellow and dark striations from bottom left to top right throughout the entire transit duration shows the stellar pulsation, the bright yellow strip extending from bottom right to top left is caused by the RM and CLV effects.}
    \label{fig:fft result}
\end{figure}

To disentangle the planetary signal from the the RM and CLV effects, as well as stellar pulsations, we adopted the methodology used in \citet{Johnson_2015} and \citet{Temple_2017}. This approach, originally developed for separating the components of stellar pulsations from planetary transits via Doppler tomography through Fourier transform, allows us to resolve and mitigate different types of signals. The results of the CCF map were then transformed into the frequency domain using Fourier transform, demonstrating orthogonality with respect to the original spatial configuration in the time domain. Figure~\ref{fig:fft} displays the transformed CCF map, which reveals a distinct feature extending from the bottom left to the top right, attributed to the CLV and RM effects, alongside structures resulting from stellar pulsations and the planetary atmosphere. These structures are aligned with the horizontal axis and distributed along nearly diagonal directions that extend from the bottom right to the top left. Subsequently, we constructed a Fourier filter that assigns a value of zero to the region containing the RM and CLV effects as well as stellar pulsations, while maintaining a value of one in the regions corresponding to the planetary signal. The Hann function was applied to mitigate discontinuities in the transition region. Finally, we performed the inverse Fourier transform on the filtered CCF map, expecting the planetary signal to manifest at the planet trace within the CCF map after the Fourier transformation, with most components of the RM and CLV effects and stellar pulsations already removed. The CCF map and the model, which discriminates frequency components related to RM and CLV effects as well as stellar pulsations, are illustrated in Figure~\ref{fig:fft result}. We employed this Fourier transform method for all planets to eliminate prominent structures in the CCF map and search for potential planetary signals.

\begin{figure}
	\includegraphics[width=8cm, height=6cm]{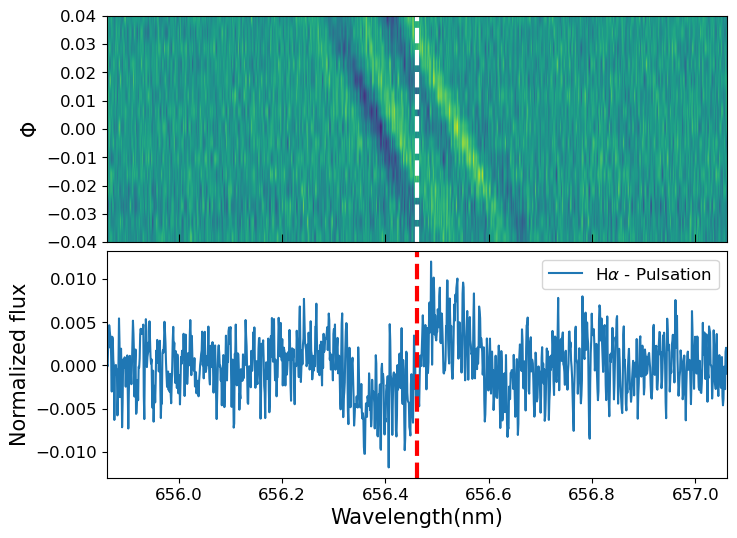}
    \caption{The two-dimensional residuals map only containing stellar pulsation after employing LSD and fourier transform method \textit{Top panel}, the white  dotted lines indicates the rest frame. The transmission spectrum of stellar pulsation is derived by summing two-dimensional residuals in transit and scaling the depth between mean line profiles and absorption lines \textit{Bottom panel}, the red dotted lines line indicates the rest frame.}
    \label{fig:pulasation_spectrum}
\end{figure}

Additionally, it is also necessary to mitigate the impact of stellar pulsation on the extracted planetary transmission spectrum. We employed the technique described in \citet{Borsa_2021} to compute the stellar mean line profiles at each orbital phase using the least-squares deconvolution (LSD) algorithm. The residuals of mean line profile are obtained after subtracting the contribution of out-of-transit mean line profile, and we then extracted the components of stellar pulsation after employing the Fourier transform method. The transmission
spectrum of stellar pulsation is derived after shifting all in-transit residuals of mean line profile to the planetary rest frame and re-scaling the ratio based on the depth of mean line profile and absorption lines at corresponding position. We derived the stellar pulsation spectrum for different species and mitigate them to obtain the cleaner planetary transmission spectra as shown in~\ref{fig:pulasation_spectrum}.

%%%%%%%%%%%%%%%%%%%%%%%%%%%%%%%%%%%%%%%%%%%%%%%%%%%%%
%%%%%%%%%%%%%%%%%%%%%%%%%%%%%%%%%%%%%%%%%%%%%%%%%%%%%

%%%%%%%%%%%%%%%%%%%%%%%%%%%%%%%%%%%%%%%%%%%%%%%%%%%%%
%%%%%%%%%%%%%%%%%%%%%%%%%%%%%%%%%%%%%%%%%%%%%%%%%%%%%

%%%%%%%%%%%%%%%%%%%%%%%%%%%%%%%%%%%%%%%%%%%%%%%%%%%%%
%%%%%%%%%%%%%%%%%%%%%%%%%%%%%%%%%%%%%%%%%%%%%%%%%%%%%

\subsection{Atmospheric elemental abundance}
\citet{Brogi_2019} first applied the Bayesian atmospheric retrieval framework to high-resolution spectroscopy, providing a unique method for placing quantitative constraints on the structure of planetary temperature-pressure profiles and atmospheric abundances. In this study, we adopted the formula further developed by \citet{Gibson_2022} to establish a complete Gaussian likelihood function for our datasets. We consider only the noise scale factor ($\beta$), while the model scale factor ($\alpha$) is set to one to reduce additional degeneracy introduced by correlated parameters~\citep{Boucher_2023}. The formula is as follows:

%--------------------------------------%
\begin{align}
  \mathcal{L}(\boldsymbol{\theta}) = \prod_{i=1}^{N} \frac{1}{\sqrt{2\pi(\beta\sigma_i)^2}} \exp\bigg(-\frac{1}{2}\frac{(f_i - m_i)^2}{\beta\sigma_i^2}\bigg)
 \label{likelihood_raw}
\end{align}
%--------------------------------------%
Where $f_{i}$ is the observation data, $\sigma_i$ is the corresponding uncertainties, $m_{i}$ is the atmospheric template containing the model parameters, N is the number of observation data for each exposure and $i$ refers to the wavelength sequence, $\beta$ is the noise scale factor. It is more intuitive to employ the log-likelihood instead of the natural logarithm of likelihood, using the following formula for the log-likelihood.

\begin{align}
    \ln{\mathcal{L}} = -\frac{N}{2}\ln{2\pi} - \sum_{i=1}^{N}\ln{\sigma_i} - N\ln{\beta} - \frac{1}{2}\chi^2
        \label{likelihood_log}
\end{align}
Where,
\begin{align}
        \chi^2 = \sum_{i=1}^{N}\frac{(f_i - m_i)^2}{(\beta\sigma_i)^2}
        \label{likelihood_chi2}
\end{align}
By expanding equation(\ref{likelihood_chi2}), we obtains,
\begin{align}
    \label{likelihood_chi2_all}
    \chi^2 = \frac{1}{\beta^2}\Bigg(\sum_{i=1}^{N}\frac{f_i^2}{\sigma_i^2} + \sum_{i=1}^{N}\frac{m_i^2}{\sigma_i^2} -2\sum_{i=1}^{N}\frac{f_im_i}{\sigma_i^2}\Bigg)
\end{align}
The log-likelihood value primarily depends on the correlation between the observational data and the established forward model, as indicated in equation (\ref{likelihood_chi2}), which is analogous to equation (\ref{ccf_formula}) in section \ref{sec:ccf}, given that other terms are constant. We utilized the template generated by \texttt{petitRADTRANS}, which includes the opacities of  Mg, Ca, Fe, Ti, and V, representing the atomic components most frequently detected in exoplanet atmospheres. The volume mixing ratios (VMRs) of these atoms are treated as free parameters to be retrieved and assigned uniform priors to determine the maximum probability regions. Additionally, we incorporated the equilibrium temperature ($T_{eq}$), the ratio between infrared and optical opacities ($K_{IR}$), the Guillot parameter ($\gamma$) and the reference surface gravity ($g$) as input parameters to construct a Guillot pressure-temperature profile. The reference surface gravity ($g$) were calculated and fixed using the planet mass and radius provided in the Table~\ref{obs}. $T_{eq}$, $K_{IR}$ and $\gamma$ are also treated as free parameters to be retrieved.

In contrast to the visual inspection of strong absorption lines in transmission spectra and the cross-correlation method used to determine the presence of atoms and molecules with multiple spectral lines in the planetary atmosphere, the likelihood function allows for a comprehensive exploration of planetary parameter space. It also imposes upper limits on the volume mixing ratio of each species by characterizing the region of probability density associated with the distribution of maximum likelihood values.

\section{RESULTS AND DISCUSSION}
\label{sec:results and discussion}
In this section, we present the results obtained by inspecting individual lines in the transmission spectra around the Ca~{\sc ii} H\&K, Na~{\sc i} D1$\&$D2, H$\alpha$, H$\beta$, Mg~{\sc i}, Li~{\sc i} and Fe~{\sc i}, we also present the searched results for these five exoplanets using the CCF technique between the transmission spectra with telluric contamination corrected and the atmospheric template spectra. The temperature structure and elemental abundance profiles of Mg, Ca, Fe, Ti and V elements is also retrieved by applying the Bayesian retrieval framework to high-resolution spectroscopy.

 \begin{table*}
      \caption{Summary of the derived parameters of the atomic lines from transmission spectrum}
         \label{tab:atomic_lines}
         %\centering
         \begin{tabular}{lcccccc}
            \hline
            \hline
            \noalign{\smallskip}
            Planets&Species & $\lambda$ (nm) &h (\%)&$V_{\rm wind}$ (km s$^{-1}$)& FWHM (km s$^{-1}$)&R$_{\lambda}$ ($R_{\rm p}$)\\
            \noalign{\smallskip}
            \hline
            WASP-50\,b&Na~{\sc i} &589.756/589.158 &0.51\,$\pm\,0.19$  & -0.96\,$\pm\,7.19$&22.91\,$\pm\,15.59$& 1.13\,$\pm\,0.09$\\
            \hline
            WASP-117\,b&H$\alpha$ &656.461 &0.51\,$\pm\,0.16$ &-3.4$\pm\,3.19$&21.39\,$\pm\,7.14$&1.29\,$\pm\,0.16$\\
            &Li~{\sc i}&670.961 &0.25\,$\pm\,0.08$  & -1.27\,$\pm\,4.04$&25.01\,$\pm\,9.39$& 1.15\,$\pm\,0.09$\\
            \hline
           &Na~{\sc i} D1 &589.756 &0.27\,$\pm\,0.23$  & -1.49\,$\pm\,17.69$&44.32\,$\pm\,40.9$& 1.17\,$\pm\,0.25$\\
           &Na~{\sc i} D2 &589.158 &1.3\,$\pm\,0.81$  & -0.87\,$\pm\,2.44$&5.08\,$\pm\,5.22$& 1.64\,$\pm\,0.64$\\
            &Ca~{\sc ii} H  &396.959 &1.9\,$\pm\,0.8$  & -52.67\,$\pm\,21.98$&104.21\,$\pm\,49.28$&1.87\,$\pm\,0.56$\\
            WASP-167\,b&H$\alpha$ &656.461 &2.41\,$\pm\,0.52$  & -1.11\,$\pm\,2.95$&27.91\,$\pm\,6.79$&2.03\,$\pm\,0.33$\\
            &Mg~{\sc i}&518.505 &1.02\,$\pm\,0.4$  & -4.38\,$\pm\,2.65$&13.6\,$\pm\,6.06$& 1.53\,$\pm\,0.34$\\
            &Fe~{\sc i}&404.695 &1.55\,$\pm\,0.46$  & -17.76\,$\pm\,6.37$&45.05\,$\pm\, 14.65$& 1.74\,$\pm\,0.34$\\
            \hline
            WASP-173A\,b&H$\alpha$ &656.461 &1.04\,$\pm\,0.5$  & -29.82\,$\pm\,2.14$&8.28\,$\pm\,4.05$&1.33\,$\pm\,0.27$\\
            \hline
            \noalign{\smallskip}
         \end{tabular}
         
         \textbf{Notes}:The line center wavelength is in vacuum, the depth of absorption line $h$, the Doppler shift of line center $V_{\rm wind}$, line width (FWHM) and the effective planetary radius, $R_{\lambda}$.
 
     \end{table*}
% %--------------------------------------------------%

\subsection{WASP-50b}
\begin{figure*}
  \centering
  \includegraphics[width=17cm]{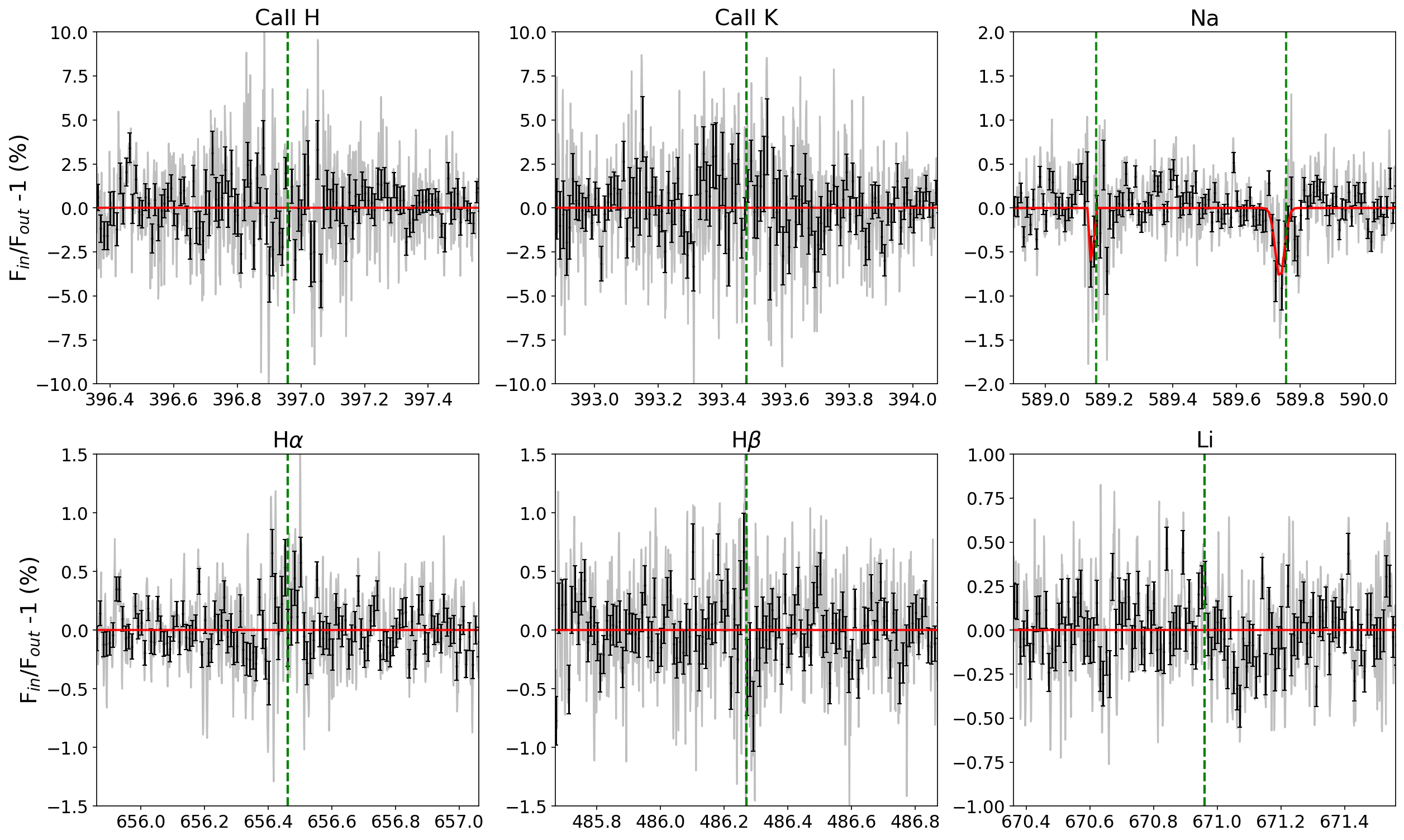}
  \caption{The 1D transmission spectra of WASP-50b around Ca~{\sc ii} H \& K, Na doublet lines, H$\alpha$, H$\beta$, and Li lines, which have applied the correction for RM and CLV effects. The integrated transmission spectra in PRF are shown in grey (original), black points with error bars (binned), and the best Gaussian fits are shown in red profiles; the horizontal red lines at zero indicate no significant detection for this species. The dashed green vertical line represents the static position of each line at the vacuum wavelength.}
  \label{fig:w50}
\end{figure*}

As shown in Fig~\ref{fig:w50}, we report the additional absorption of Na~{\sc i} with the significance levels of $\sim$2.7$\sigma$ and blueshifted veloctiy of -0.96 km\,s$^{-1}$ as measured from the reference position, the measurement of Na absorption is $\sim$0.51\%, which corresponds to the effective planetary radius of 1.13 $R_{p}$. No significant absorption signal is detected for any other atmospheric species in transmission spectra and CCF maps.

\subsection{WASP-117b}
The searched results in transmission spectra for WASP-117\,b is shown in Fig~\ref{fig:w117}. We report a tentative detection of the H$\alpha$ and Li~{\sc i} lines as well with significance levels of $\sim$3.2$\sigma$ and $\sim$3.1$\sigma$, respectively, as detailed in Table~\ref{tab:atomic_lines}, we are unable to confirm or reject these spectral signals as originating from the planet based on the current data set, the notable fluctuations and errors around the lines of Ca~{\sc ii} H\&K, H$\beta$ and Fe~{\sc i} also suggest the observational data may be extensively influenced by the stellar processes. Additionally, we exclude the analysis of Na~{\sc i} due to the presence of anomalous flux of the stellar spectra around Na doublet lines, which resulted in anomalous absorption features.  The thorough analysis of the CCF maps revealed no significant absorption features attributed to the possible atmospheric species. More data is needed to confirm the tentative detection reported here. Additional observations are necessary to constrain further and understand the atmosphere of WASP-117\,b.

\subsection{WASP-156b}
\begin{figure*}
  \centering
  \includegraphics[width=17cm]{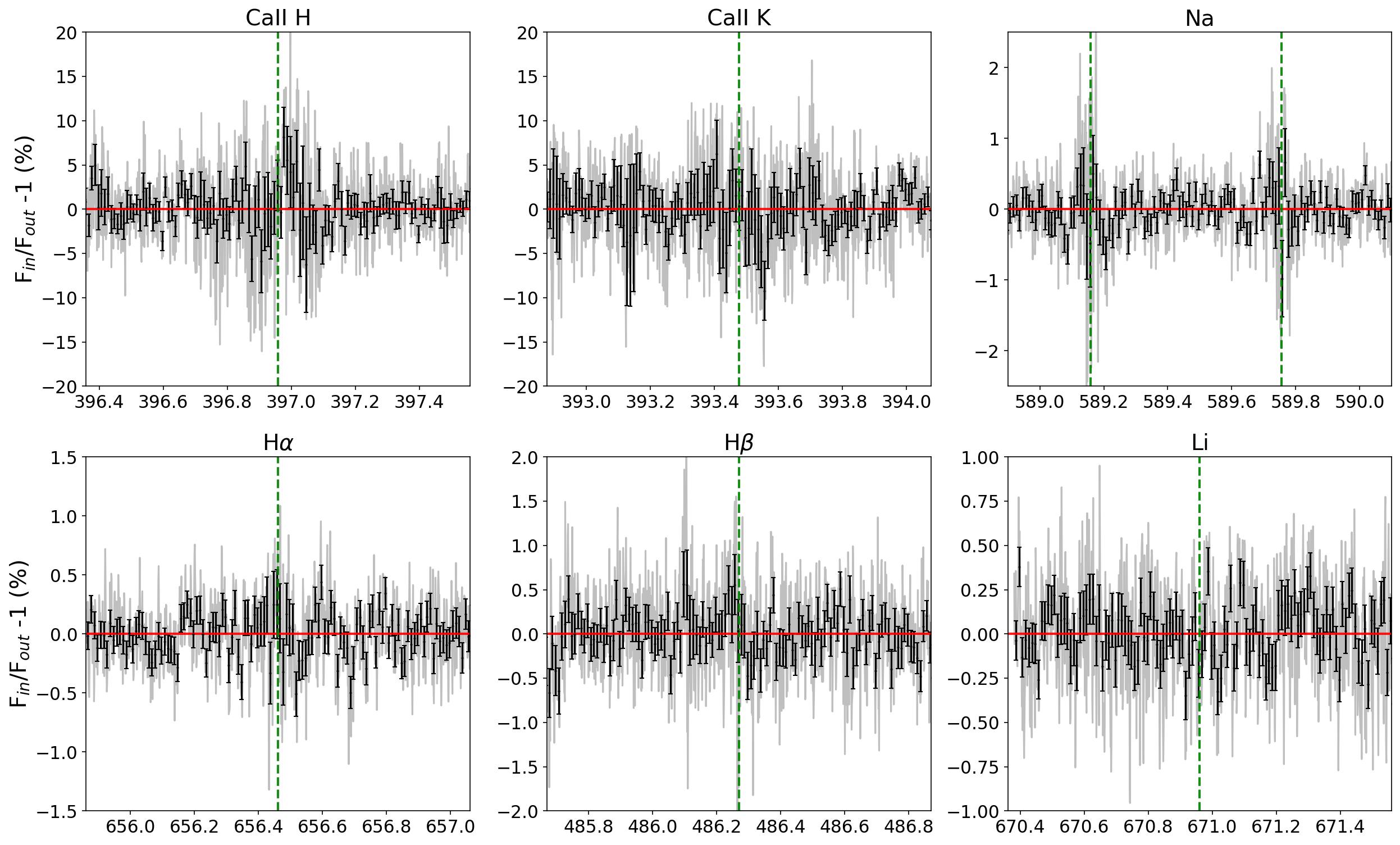}
  \caption{Same as Fig~\ref{fig:w50}, but for WASP-156b}
  \label{fig:w156}
\end{figure*}

Absorption signals for any other atmospheric species were absent in both the transmission spectra, as shown in Fig~\ref{fig:w156}, and CCF maps analyzed from the current observational data. ~\citet{JiangCZ_2023} also found a featureless feature in the transmission spectra with the large uncertainties of WASP-156\,b using the low-resolution spectra observed by Optical System for Imaging and low-intermediate Resolution Integrated Spectroscopy on the Gran Telescopio Canarias (GTC-OSIRIS) and suggested that the absence of discernible signals could not be attributed to cloudy atmospheres. Given the planet’s relatively low equilibrium temperature, possible clouds or haze layers in the upper atmosphere may suppress inherently weak spectral features, necessitating observations with higher SNR and broader wavelength coverage for conclusive atmospheric characterization.

\subsection{WASP-167b}
\begin{figure*}
  \centering
  \includegraphics[width=17cm]{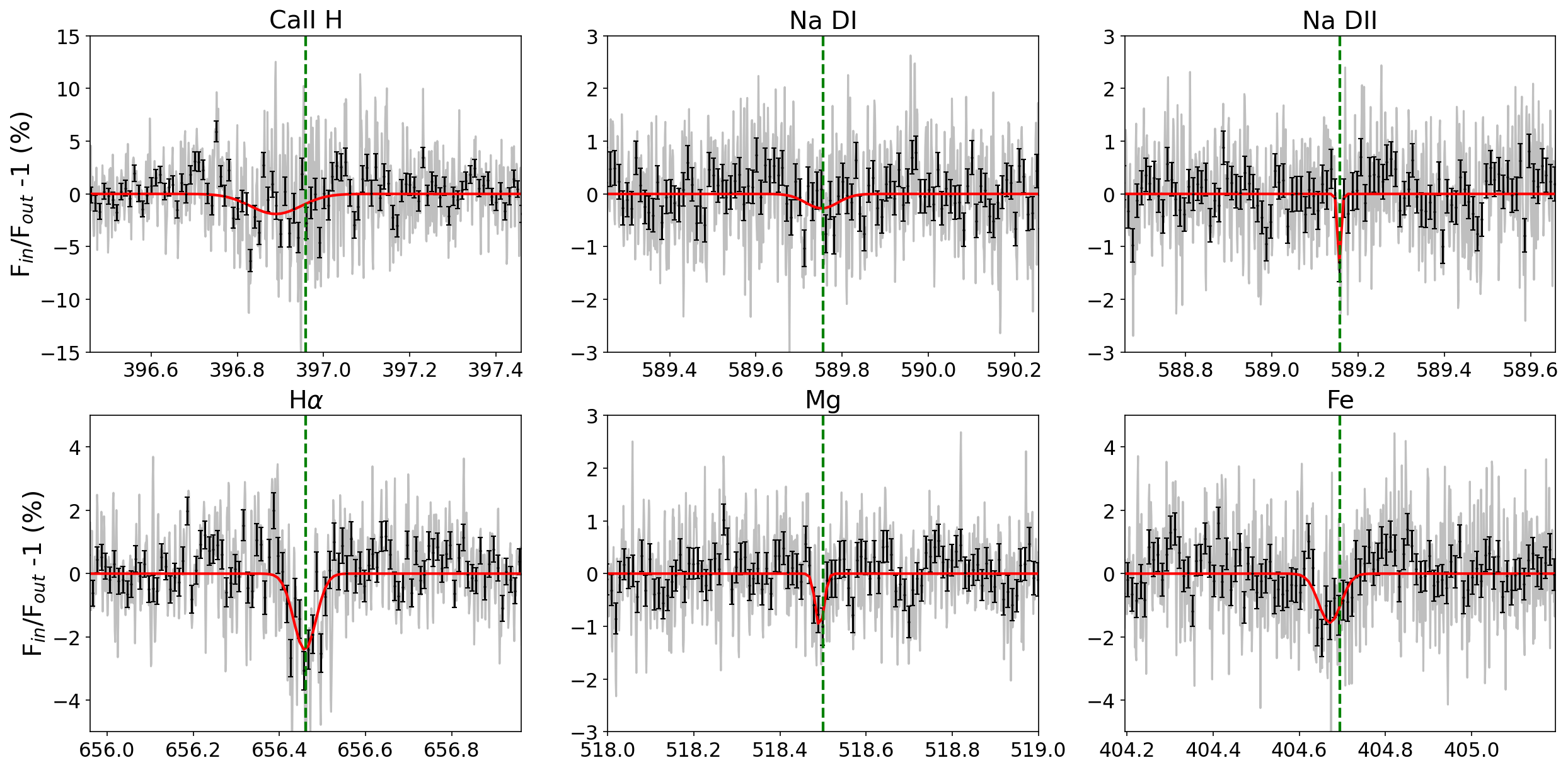}
  \caption{Same as Fig~\ref{fig:w50}, but including Mg and Fe lines and excluding Ca~{\sc ii} K and Li lines for WASP-167b}
  \label{fig:w167}
\end{figure*}

%Gaussian fits using the python module \texttt{lmfit} are applied on
We report the first detection of additional absorption signatures from H$\alpha$ in the atmosphere of the HJ WASP-167b, with significance levels of 4.6$\sigma$, respectively. In addition,  we have obtained tentative detection of Ca~{\sc ii} H, Mg~{\sc i} and Fe~{\sc i} with significance levels of 2.4$\sigma$, 2.6$\sigma$ and 3.4$\sigma$, respectively. The Na doublet has a weak signal with below 2.0$\sigma$, probably due to the low SNR at their line cores. These individual lines are shown in Fig~\ref{fig:w167} and the corresponding measurements are summarized in Table~\ref{tab:atomic_lines}. We note that all the detected planet-origin lines are blueshifted relative to their rest frame, indicative of planetary winds transcending the terminator region toward the observer. Such phenomena have been detected previously in multiple planets~\citep{Kesseli_2022,Maguire_2023,Costa_Silva_2024}.

\begin{figure*}
    \centering
    \subfigure
    {
    \begin{minipage}[t]{1\linewidth}
    \centering
    \includegraphics[width=17.5cm,height=3.4cm]{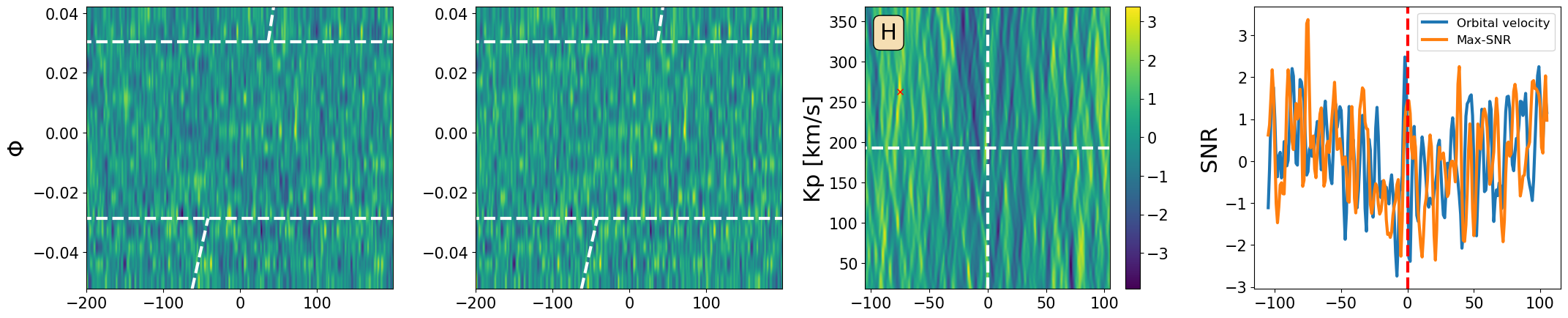}  
    \end{minipage}
    }
    \subfigure
    {
    \begin{minipage}[t]{1\linewidth}
    \centering
    \includegraphics[width=17.5cm,height=3.4cm]{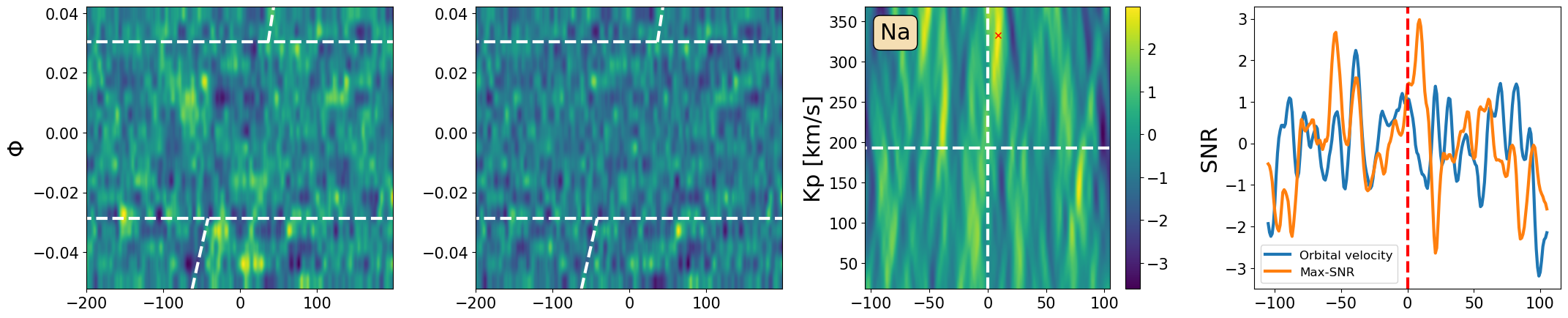}  
    \end{minipage}
    }
    \subfigure
    {
    \begin{minipage}[t]{1\linewidth}
    \centering
    \includegraphics[width=17.5cm,height=3.4cm]{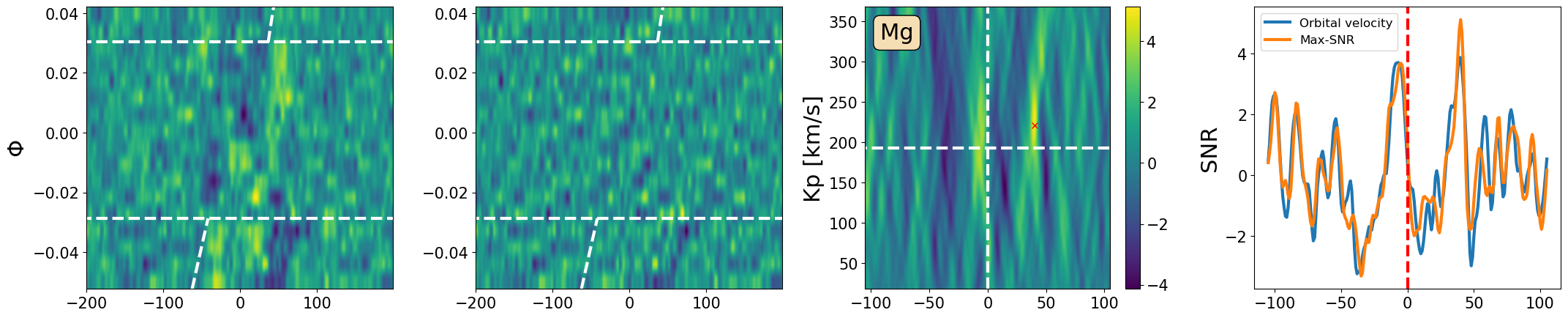}  
    \end{minipage}
    }
    \subfigure
    {
    \begin{minipage}[t]{1\linewidth}
    \centering
    \includegraphics[width=17.5cm,height=3.4cm]{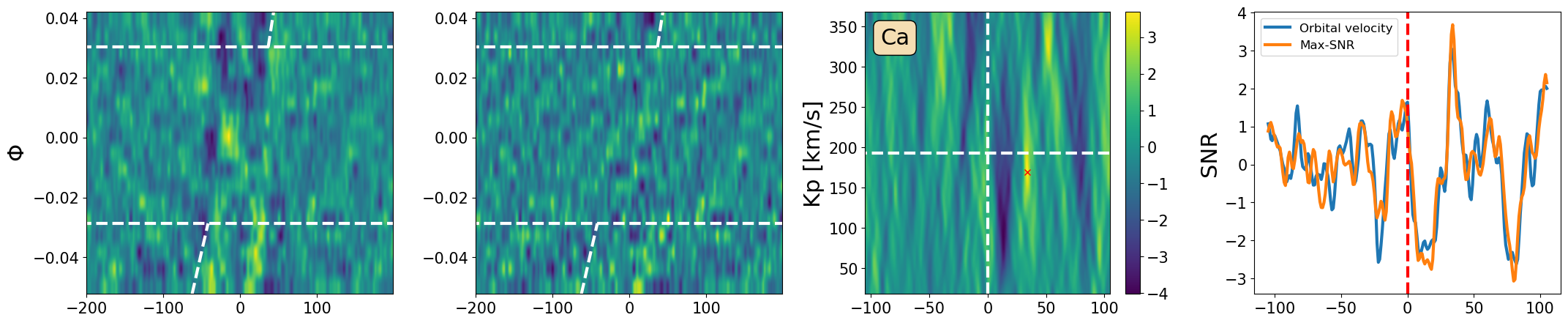}  
    \end{minipage}
    }
    \subfigure
    {
    \begin{minipage}[t]{1\linewidth}
    \centering
    \includegraphics[width=17.5cm,height=3.4cm]{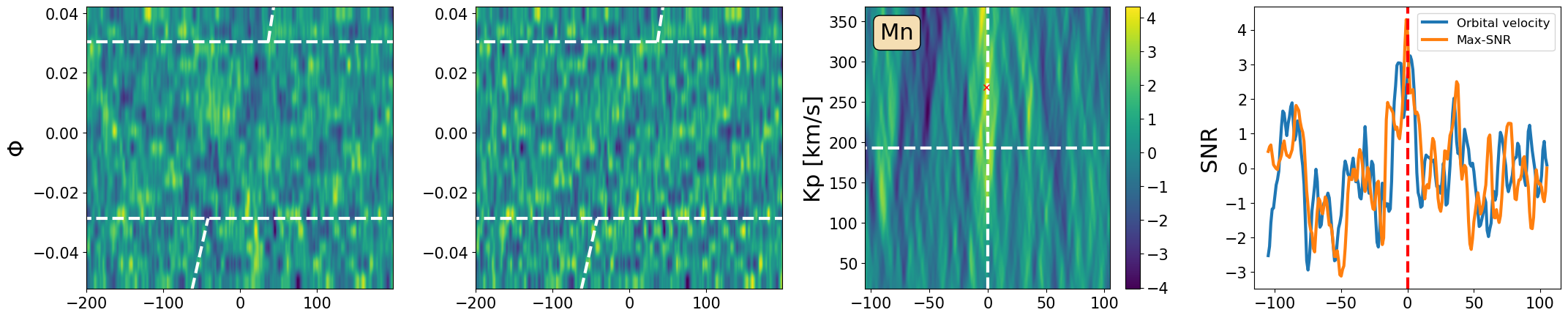}  
    \end{minipage}
    }
    \subfigure
    {
    \begin{minipage}[t]{1\linewidth}
    \centering
    \includegraphics[width=17.5cm,height=3.4cm]{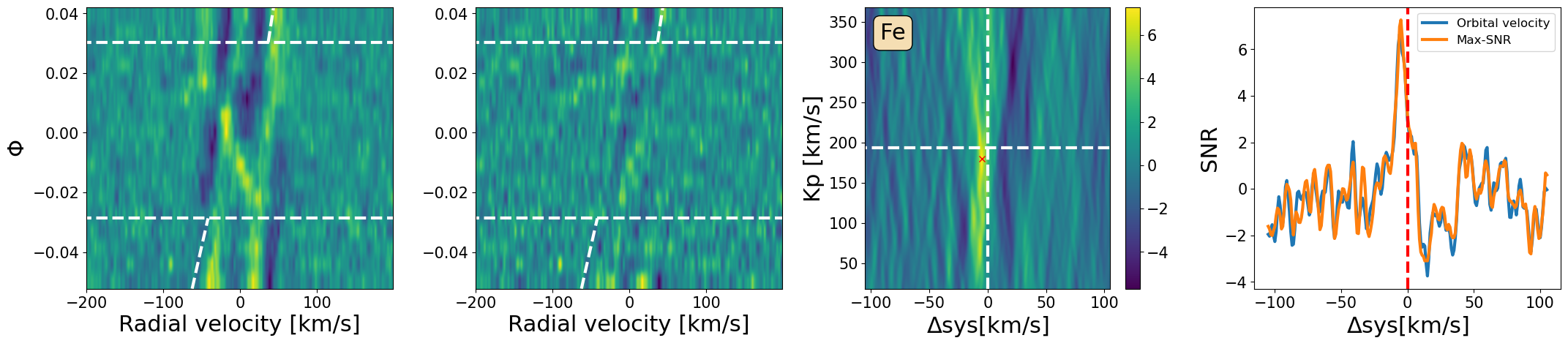}  
    \end{minipage}
    }
    
\caption{\emph{First panels}: The 2D CCF maps of H, Na, Mg, Ca, Mn and Fe for WASP-167b with RM+CLV effects and stellar pulsations uncorrected. The white dotted lines indicate the beginning and ending positions of the transit and the inclined white lines indicate the expected trace of signal from the planet. \emph{Second panels}: Same as \emph{the first panels} but with RM+CLV effects and  stellar pulsations corrected. \emph{Third panels}: the $K_{\rm p}$-$\Delta V_{\rm sys}$ maps in the range of -$20\sim350$\, km\,s$^{-1}$. In each panel, the planet signal is expected to appear around the intersection of the two white dotted lines, while the red crosses marks the position with maximum SNR. \emph{Fourth panels}: the SNR plots at the expected $K_{\rm p}$ position in blue and at the Max-SNR position in orange.}
\label{fig:ccf_result}
\end{figure*}

Furthermore, we have searched for the species from Li to U using the CCF technique, and the results are presented in Fig.~\ref{fig:ccf_result}. In this figure, the 1st and 2nd column panels show the 2D CCF maps for each species before and after correcting the RM+CLV effects and stellar pulsations. The 3rd column panels illustrate the distribution of expected signals across the domains of Keplerian velocity of the planet ($K_{\rm p}$) and the velocity relative to the planet rest frame ($\Delta V_{\rm sys}$), in which a true signal would coincide with the intersection point between two dashed lines, and measured as the significance level at different position in the fourth panel. The red crosses in the 3rd column panels mark the positions with maximum SNR, which, however, do not always coincide well with the intersection point. Gaussian fits using the Python module \texttt{lmfit} are applied on the horizontal and vertical slices passing the red cross to determine the peak positions in $\Delta V_{\rm sys}$ and $K_{\rm p}$, respectively, as well as the FWHM of the detected signals.

From our CCF analysis, the atom Fe~{\sc i} is clearly detected with a max SNR of 7.3. A Gaussian fit yields $\Delta V_{\rm sys}$ of $-5.0 \pm\,1.45$\,km\,s$^{-1}$, $K_{\rm p}$ of 180.0 $\pm\,18.41$ km s$^{-1}$ and FWHM of 10.55 $\pm\,1.02$ km s$^{-1}$ at the peak SNR in the $\Delta V_{\rm sys}-K_{\rm p}$ map, suggesting that the Fe absorption likely arises from the movements of terminal winds. The difference between the measured $\Delta V_{\rm sys}$ from the single strong line method and CCF method for Fe~{\sc i} absorption signal, $-17.76\pm6.37$ km\,s$^{-1}$ against $-5.0 \pm\,1.45$ km\,s$^{-1}$, is notable and significant. It is likely due to the not fully corrected stellar activity that can distort the stellar spectra and mimic planetary signals~\citep{Seager_2025}, which can be reasonably well corrected during the CCF analysis by applying Fourier transform technique, as described in Fig.~\ref{fig:ccf_result}, however, can not be treated well in the single line method.

\begin{figure*}
	\includegraphics[width=17cm]{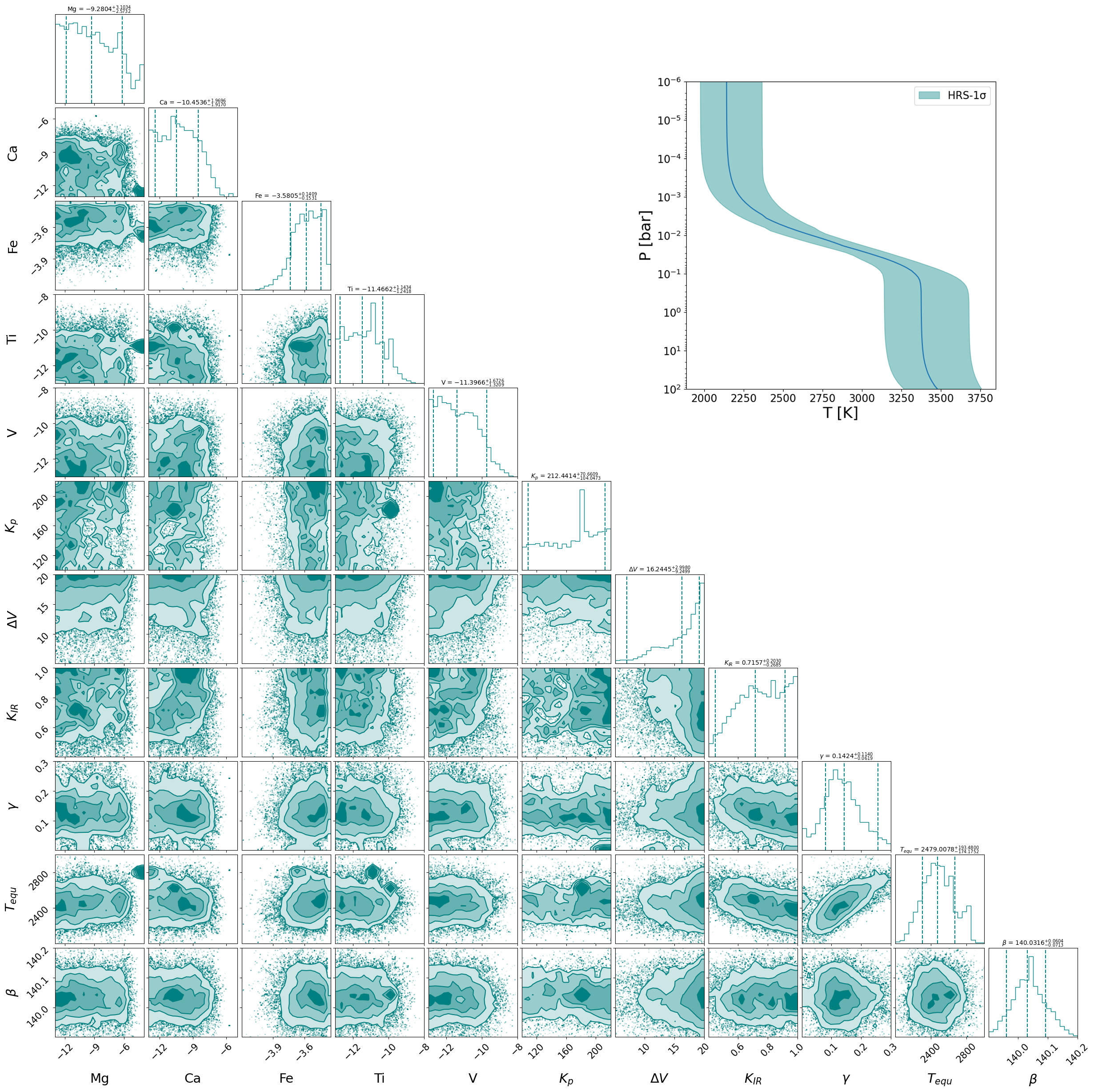}
    \caption{The posterior distributions of atmospheric parameters using the atmospheric retrieval technique based on the WASP-167b data.}
    \label{fig:W167_retrieval_results}
\end{figure*}

Additionally, atmospheric retrieval analysis using a Bayesian framework is conducted to derive the VMRs of several species, which give logarithmic VMRs of Fe, Ca, Mg, Ti and V of $-3.58^{+0.14}_{-0.15}$, $-10.45^{+1.97}_{-1.92}$, $-9.28^{+3.10}_{-2.57}$, $-11.47^{+1.14}_{-1.24}$, and $-11.40^{+1.67}_{-1.32}$, respectively. We also derive the estimated orbital velocity ($K{p}$) = $212.44^{+70.66}_{-104.05}$ km s$^{-1}$, the system velocity ($\Delta V$) = $16.24^{+3.00}_{-9.25}$ km s$^{-1}$, and the equilibrium temperature ($T_{\mathrm{equ}}$) = $2479^{+193}_{-174}$ K. The retrieval results are shown in Fig.~\ref{fig:W167_retrieval_results}. 

The results reveal a significant accumulation of Fe, consistent with the CCF results (Fig.~\ref{fig:ccf_result}), confirming it as the most prominent species detected in the atmosphere of WASP-167b. The low VMRs of the other species indicate inherently low abundances, resulting in weak signals that are likely overwhelmed by noise (e.g., stellar pulsations) and are therefore nearly undetectable. Furthermore, unlike the CCF technique, the Bayesian retrieval does not account for signals such as the RM effect. This leads to inconsistencies between the velocity-space signals derived from the two methods. It should be noted that transmission spectra affected by uncorrected stellar pulsations limit the final accuracy of the retrieval results. Consequently, accurately determining the parameters of planetary atmospheric species affected by stellar activity remains challenging.

\subsection{WASP-173A\,b}
\begin{figure*}
  \centering
  \includegraphics[width=17cm]{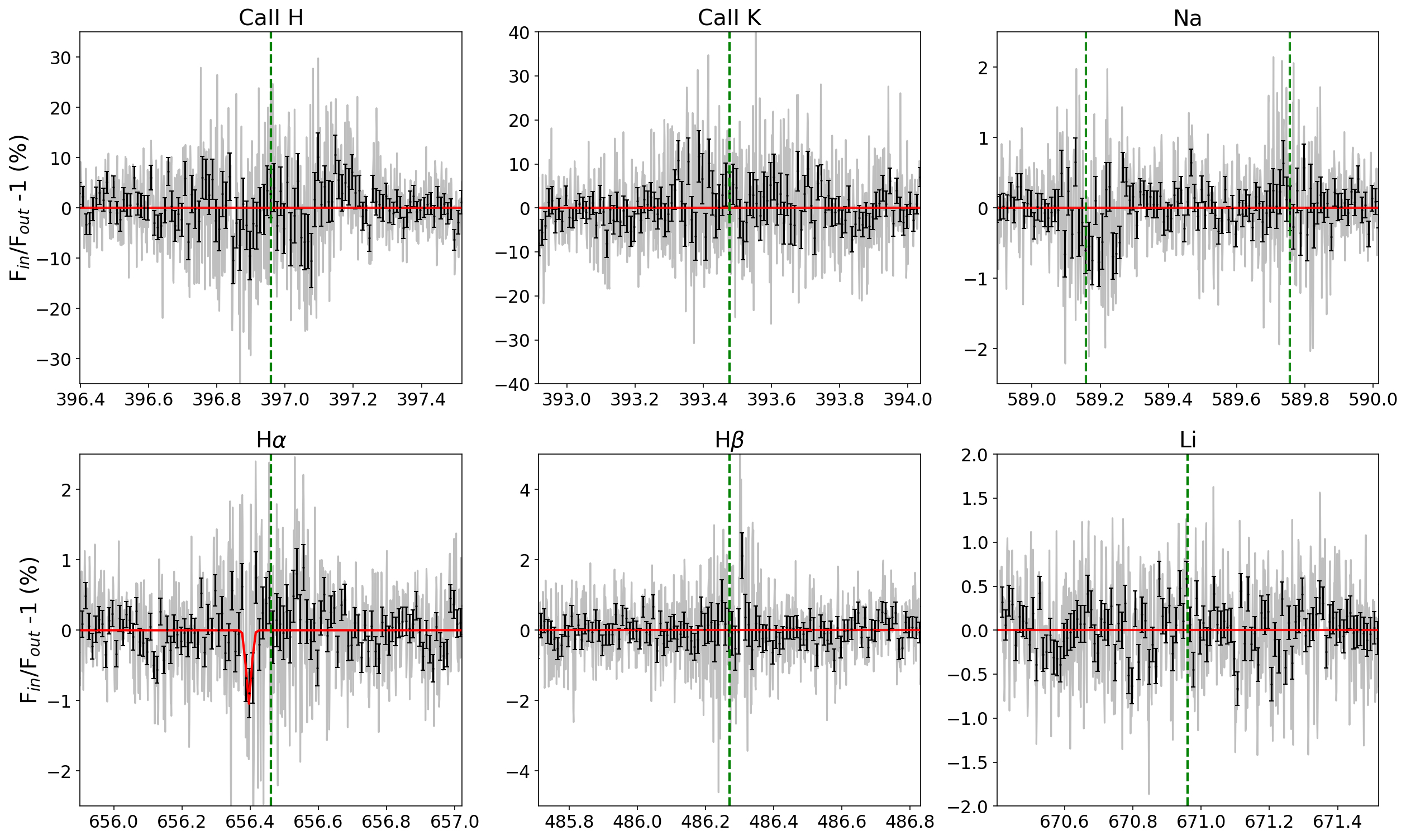}
  \caption{Same as Fig~\ref{fig:w50}, but for WASP-173A\,b}
  \label{fig:w173}
\end{figure*}

WASP-173A\,b, as a planet orbiting one star in a binary system, is exposed to more intense stellar radiation and influenced by more complex gravitational forces, providing a useful lab for the studies of the atmospheric composition, thermal structure, and dynamics of exoplanets under complex conditions~\citep{Johnstone_2019}. As shown in Fig~\ref{fig:w173}, we report a marginal additional absorption of H$\alpha$ of $\sim1.04\pm0.5$\%, corresponding to an effective planetary radius of 1.33 $R_{p}$ at a blue-shifted velocity of $-29.82$km\,s$^{-1}$ with respect to its host star. No significant absorption signal is detected in transmission spectra or CCF maps. 

\begin{figure*}[htbp]%refree:6cm
  \centering
  \includegraphics[width=5.5cm]{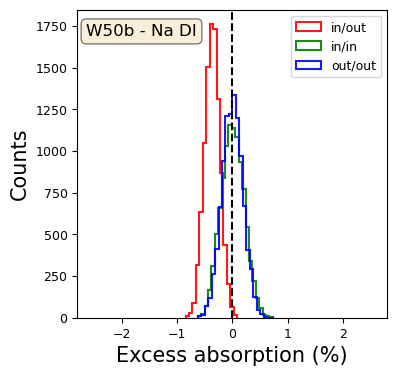}
  \includegraphics[width=5.5cm]{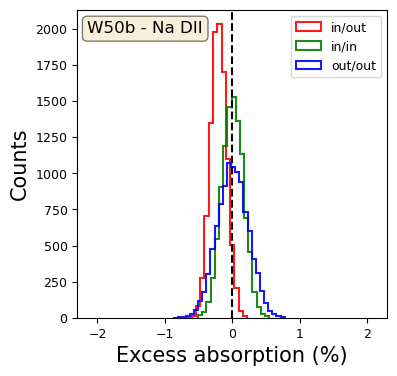}
  \includegraphics[width=5.5cm]{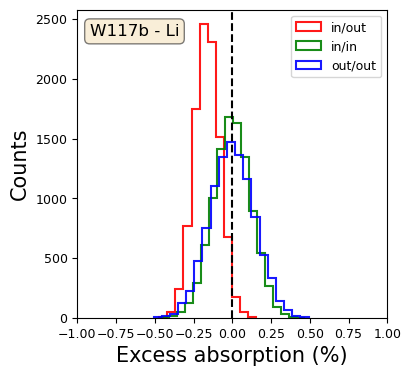}
  \includegraphics[width=5.5cm]{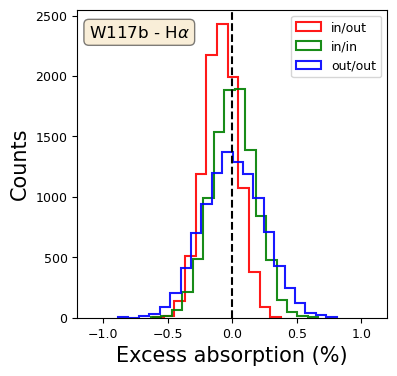}
  \includegraphics[width=5.5cm]{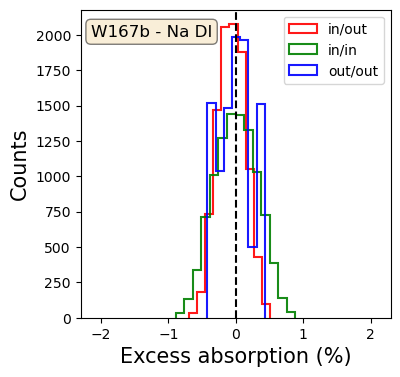}
  \includegraphics[width=5.5cm]{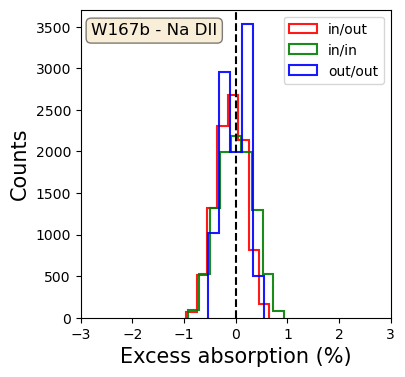}
  \includegraphics[width=5.5cm]{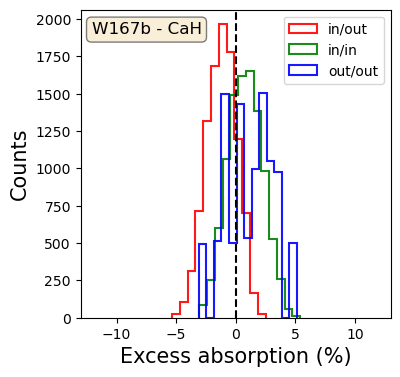}
  \includegraphics[width=5.5cm]{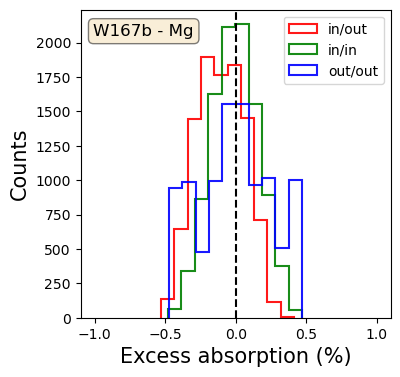}
  \includegraphics[width=5.5cm]{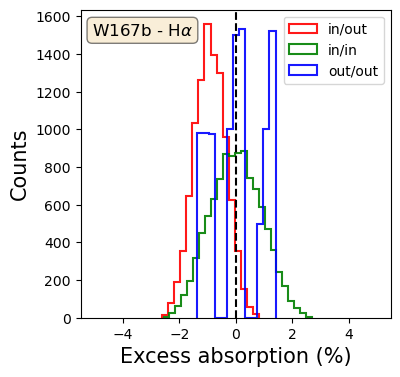}
  \includegraphics[width=5.5cm]{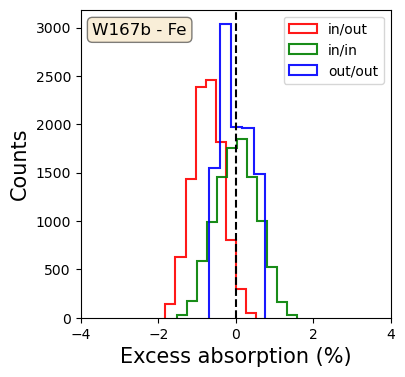}
  \includegraphics[width=5.5cm]{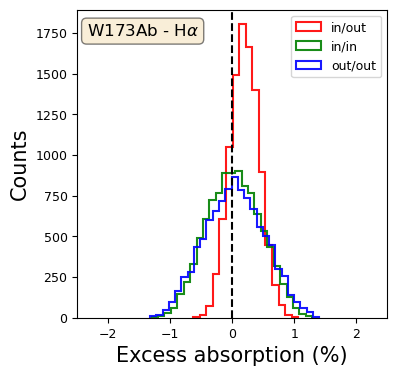}
  \caption{Results of EMC simulation for the species in the atmosphere of these exoplanets that may be detected as listed in Table~\ref{tab:atomic_lines}. These distributions for three EMC scenarios are shown in different colors: red is the in-out distribution; green is the in-in distribution; and blue is the out-out distribution.}
  \label{fig:emc for atom}
\end{figure*}

Additionally, the results derived by EMC simulations are shown in Fig.~\ref{fig:emc for atom}. The in-out distributions of  Na~{\sc i} in WASP-50b, Li~{\sc i} and H$\alpha$ in WASP-117b, H$\alpha$ and Fe~{\sc i} in WASP-167b exhibit clear excess absorption, while the absorption depths in the in-in and out-out scenarios are centered at zero. The EMC simulations suggest that these signals are likely to have been created by the transits and may originate from the planet itself. The distributions of Mg~{\sc i} and Na~{\sc i} in WASP-167b exhibit a subtle excess absorption but remain poorly resolved from the other two scenarios. This may be attributed to the inherent weakness of the Na within the atmosphere of WASP-167b. The distributions of CaH lines in WASP-167b also exhibit some excess absorption, while the absorption depths in the in-in and out-out scenarios are not centered at zero. This may be attributed to the intense stellar activity, which precludes the definitive confirmation of Ca~H absorption in this planet's atmosphere. Note that the distribution of the out-of-transit scenario in WASP-167b exhibits a large distribution spread, possibly due to the limited sample size of ($\sim$ 6) the out-of-transit spectra and thus is more susceptible to stellar pulsation. Finally, the distribution of H$\alpha$ in WASP-173Ab shows a subtle excess emission, which cannot be robustly confirmed to confirm the existence of a signal in the atmosphere of WASP-173Ab.

\begin{figure*}
  \centering
  \includegraphics[width=17cm]{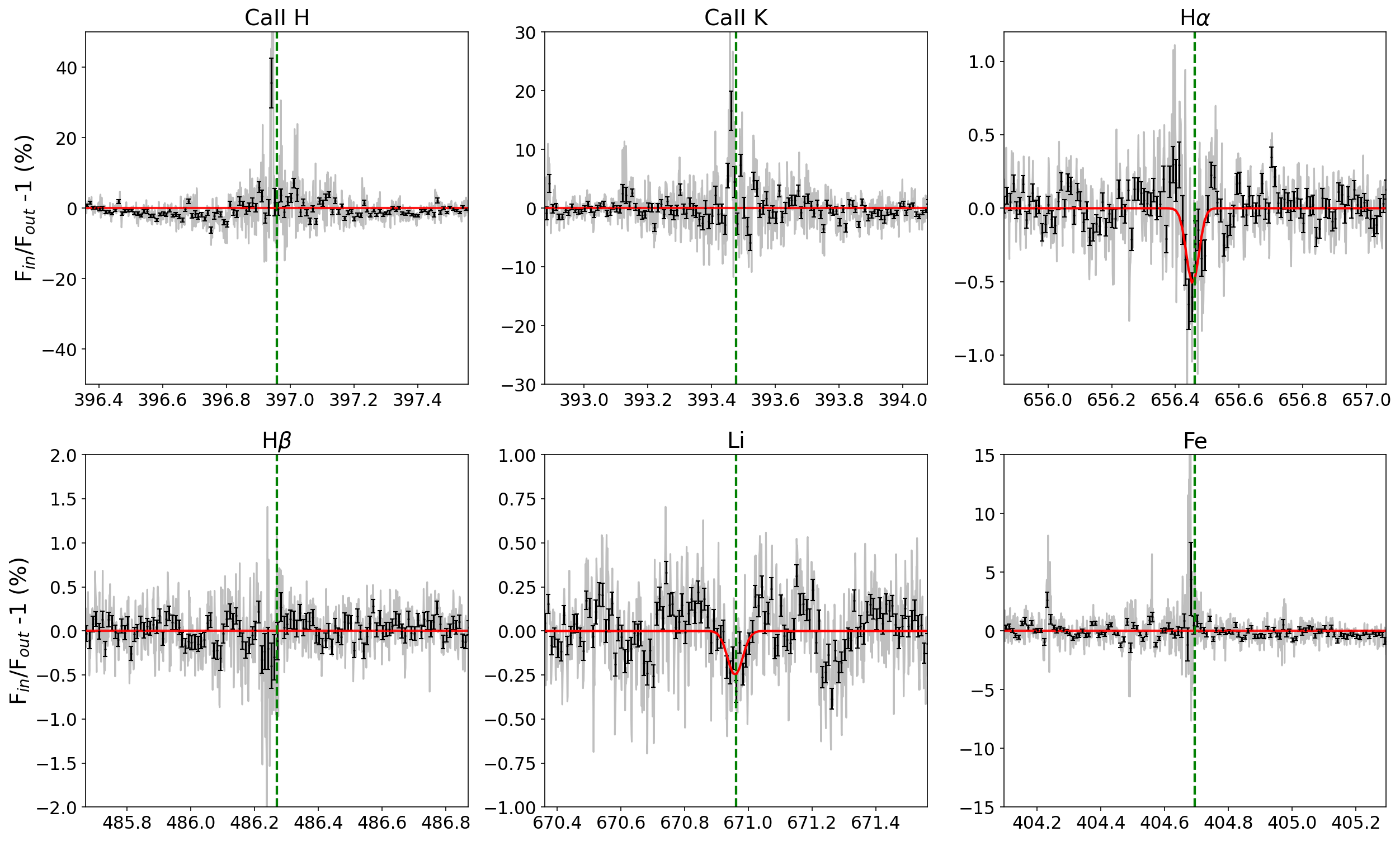}
  \caption{Same as Fig~\ref{fig:w50}, but including Fe line and excluding Na doublet lines for WASP-117b}
  \label{fig:w117}
\end{figure*}

As expected, the strength of the planetary atmospheric signal exhibits some positive correlation with equilibrium temperature. Although WASP-167b is strongly affected by its highly active host star, its atmosphere still retains a relatively rich chemical composition as presented in Table~\ref{tab:atomic_lines}. Moreover, WASP-167b exhibits the highest planetary equilibrium temperature and the lowest surface gravity of ~3.5 m s$^{-2}$ as shown in the bottom panel of Figure~\ref{fig:PM_relationship}, enhancing its atmospheric detectability. Although WASP-173Ab retains a higher equilibrium temperature, it also exhibits the highest surface gravity of $\sim$52\,m\,s$^{-2}$ among these samples, reaching nearly twice that of Jupiter. This may suggest that its atmospheric constituents are constrained by a very strong gravitational field, resulting in low detectability. WASP-156b with a mass of about 0.128\,$M_{\rm J}$ and a radius of 0.51 $R_{\rm J}$, nearly classified as a Neptune-like planet, resides at the boundary between the Hot Jupiter and Neptune-like planet populations, which provides us with an opportunity to investigate the evolution of atmospheric constituents across this region. While the non-detection of planet signals in the atmosphere of WASP-156b may be caused by its relatively low equilibrium temperature, and thus the small atmospheric scale height and transmission spectroscopy metric~\citep[TSM,][]{Kempton_2018}. The results of WASP-117b are also unreliable due to the transmission spectra exhibiting significant fluctuations as shown in Fig~\ref{fig:w117}. 

We also show that removing RM effects and stellar pulsation by applying the FFT technique for a planet similar to WASP-167b is feasible and useful. This technique can be utilized as a standard methodology for removing the influence of stellar activity and the line distortion of stellar spectra induced by a transiting planet passing before a rotating star with non-uniform surface brightness. Ultimately, more transit observations are required to improve the SNR of detection results in exoplanet atmospheres, and utilizing facilities such as the James Webb Space Telescope, which can extend the wavelength coverage into the near-infrared band, will increase the range of detectable atmospheric constituents. These improvements can enable us to better understand the differing atmospheric characteristics among various types of exoplanets and also provide a deeper insight into the processes of planetary formation and evolution.

To summarize, we have performed thorough atmospheric studies of WASP-50\,b WASP-117\,b, WASP-156\,b, WASP-167\,b and WASP-173A\,b, and obtained detection or marginal detection of $H\alpha$ absorption in three hotter ones, namely WASP-117\,b, WASP-167\,b and WASP-173A\,b. In addition, we detected alkali metal absorption in WASP-50\,b, WASP-117\,b, and WASP-167\,b. The latter planet is the hottest in this sample, and, like other UHJs, it contains a significant amount of metals in its atmosphere, including Na~{\sc i}, Ca~{\sc ii}, Mg~{\sc i}, and Fe~{\sc i}. We find that, in general, planets with higher $T_{\rm eq}$ exhibit stronger atmospheric signals, which we attribute mainly to two factors. One is that a higher $T_{\rm eq}$, usually accompanied by lower density, yields a larger scale height and thus a larger absorption area. The second reason is that freeing metal atoms/ions from their solid compounds and lifting them into the upper atmosphere requires a high $T_{\rm eq}$. By contrasting WASP-167\,b and WASP-173A\,b, and other UHJs like MASCARA-4\,b~\citep{Jiang_2023a}, we confirm a transition occurring at $T_{\rm eq}$ between 1900 and 2300\,K, above which metal lines become prominent enough for meaningful detection, as pointed out by \cite{Kitzmann_2018,Snellen2025}.

%%%%%%%%%%%%%%%%%%%%%%%%%%%%%%%%%%%%%%%%%%%%%%%%%%%%%%%%
%%%%%%%%%%%%%%%%%%%%%%%%%%%%%%%%%%%%%%%%%%%%%%%%%%%%%%%%

%%%%%%%%%%%%%%%%%%%%%%%%%%%%%%%%%%%%%%%%%%%%%%%%%%%%%%%%
%%%%%%%%%%%%%%%%%%%%%%%%%%%%%%%%%%%%%%%%%%%%%%%%%%%%%%%%

%%%%%%%%%%%%%%%%%%%%%%%%%%%%%%%%%%%%%%%%%%%%%%%%%%%%%%%%
%%%%%%%%%%%%%%%%%%%%%%%%%%%%%%%%%%%%%%%%%%%%%%%%%%%%%%%

%%%%%%%%%%%%%%%%%%%%%%%%%%%%%%%%%%%%%%%%%%%%%%%%%%%%%%%%
%%%%%%%%%%%%%%%%%%%%%%%%%%%%%%%%%%%%%%%%%%%%%%%%%%%%%%%

%%%%%%%%%%%%%%%%%%%%%%%%%%%%%%%%%%%%%%%%%%%%%%%%%%%%%%%%
%%%%%%%%%%%%%%%%%%%%%%%%%%%%%%%%%%%%%%%%%%%%%%%%%%%%%%%

%%%%%%%%%%%%%%%%%%%%%%%%%%%%%%%%%%%%%%%%%%%%%%%%%%%%%%%
%%%%%%%%%%%%%%%%%%%%%%%%%%%%%%%%%%%%%%%%%%%%%%%%%%%%%%%

%---------------------------------------------------------------------------------------------------------------------------------%

%%%%%%%%%%%%%%%%%%%%%%%%%%%%%%%%%%%%%%%%%%%%%%%%%%%%%
%%%%%%%%%%%%%%%%%%%%%%%%%%%%%%%%%%%%%%%%%%%%%%%%%%%%%

%% IMPORTANT! The old "\acknowledgment" command has be depreciated. It was
%% not robust enough to handle our new dual anonymous review requirements and
%% thus been replaced with the acknowledgment environment. If you try to 
%% compile with \acknowledgment you will get an error print to the screen
%% and in the compiled pdf.
%% 
%% Also note that the akcnowlodgment environment does not support long amounts of text. If you have a lot of people and institutions to acknowledge, do not use this command. Instead, create a new 
\section{Acknowledgments}
We thank the anonymous reviewer for their constructive comments.
This research is supported by the National Key R\&D Program of China No.~2024YFA1611802, SQ2025YFE0102100, SQ2025YFE0213204, the National Natural Science Foundation of China grants No.12588202, 62127901, 12273055,  the National Astronomical Observatories Chinese Academy of Sciences No. E4TQ2101, the China Manned Space Project with NO. CMS-CSST-2025-A16 and the Pre-research project on Civil Aerospace Technologies No. D010301 funded by China National Space Administration (CNSA).

\bibliography{sample631}{}
\bibliographystyle{aasjournal}

%% This command is needed to show the entire author+affiliation list when
%% the collaboration and author truncation commands are used.  It has to
%% go at the end of the manuscript.
%\allauthors

%% Include this line if you are using the \added, \replaced, \deleted
%% commands to see a summary list of all changes at the end of the article.
%\listofchanges

\end{document}